\documentclass[12pt]{article}

\pdfoutput=1

\usepackage[margin=1in]{geometry}
\usepackage{times}
\usepackage{graphicx,psfrag,epsf}
\usepackage{enumerate}
\usepackage{url} 
\usepackage{setspace}
\newcommand{\blind}{1}

\newlength\tindent
\usepackage[utf8]{inputenc} 
\usepackage[T1]{fontenc}    
\usepackage{hyperref}       
\usepackage{url}            
\usepackage{booktabs}       
\usepackage{amsfonts}       
\usepackage{nicefrac}       
\usepackage{microtype}      
\usepackage{comment}      
\usepackage{multirow}

\usepackage{graphicx}
\usepackage{subcaption}

\usepackage{amsmath}
\usepackage{amssymb, mathtools}
\usepackage{amsfonts}
\usepackage{mathptmx}
\DeclarePairedDelimiter{\inner}{\langle}{\rangle}

\usepackage{algorithm}
\usepackage{algpseudocode}

\usepackage{tikz}
\usetikzlibrary{positioning,shapes.geometric, arrows, fit,calc}
\tikzstyle{block} = [draw, rectangle, fill=orange!50, text width=8em, text centered, minimum height=10mm, node distance=10em]
\tikzstyle{container} = [draw, rectangle, dashed, inner sep=0.7em]
\tikzstyle{arrow} = [thick,->,>=stealth]

\usepackage{amsthm}

\usepackage{color} 
\usepackage{"./macro/GrandMacros"}
\usepackage{authblk}

\begin{document}
\setlength{\abovedisplayskip}{5pt}
\setlength{\belowdisplayskip}{5pt}
\setlength{\abovedisplayshortskip}{5pt}
\setlength{\belowdisplayshortskip}{5pt}


\if1\blind
{
\title{\bf \vspace{-3ex}
A Cross-validated Ensemble Approach to\\
Robust Hypothesis Testing of Continuous Nonlinear Interactions:\\
Application to Nutrition-Environment Studies
}
\date{\vspace{-9ex}}
\author[1]{Jeremiah Zhe Liu  \thanks{zhl112@mail.harvard.edu\\
This work was supported by grants ES007142, ES016454, ES000002, ES014930, ES013744, ES017437, ES015533, ES022585 from the National Institutes of Health.  
}}
\author[2,3]{Jane Lee}
\author[2]{Pi-i Debby Lin} 
\author[4]{Linda Valeri}
\author[2,5]{David C. Christiani}
\author[2,3]{David C. Bellinger}
\author[6]{Robert O. Wright}
\author[2,3]{Maitreyi M. Mazumdar}
\author[1,2]{Brent A. Coull}
\affil[1]{Department of Biostatistics, 
Harvard T.H. Chan School of Public Health, Boston, MA, USA.}
\affil[2]{Department of Environmental Health, 
Harvard T.H. Chan School of Public Health, Boston, MA, USA.}
\affil[3]{Department of Neurology, Boston Children's Hospital, Boston, MA, USA.}
\affil[4]{Department of Biostatistics, Columbia Mailman School of Public Health, New York, New York, USA}
\affil[5]{Department of Epidemiology, 
Harvard T.H. Chan School of Public Health, Boston, MA, USA.}
\affil[6]{Department of Environmental Medicine and Public Health, 
Icahn School of Medicine, New York, NY, USA.}
\date{}

\setcounter{Maxaffil}{0}
\renewcommand\Affilfont{\itshape\small}

\maketitle
} \fi

\if0\blind
{
  \bigskip
  \bigskip
  \bigskip
\title{\bf \vspace{-3ex}
A Cross-validated Ensemble Approach to\\
Robust Hypothesis Testing of Continuous Nonlinear Interactions:\\
Application to Nutrition-Environment Studies
}
\medskip

\author{}
\date{}

\maketitle
} \fi

\thispagestyle{empty}

\newpage
{\setstretch{1.4}
\begin{abstract}
Gene-environment and nutrition-environment studies often involve testing of high-dimensional interactions between two sets of variables, each having potentially complex nonlinear main effects on an outcome. Construction of a valid and powerful hypothesis test for such an interaction is challenging, due to the difficulty in constructing an efficient and unbiased estimator for the complex, nonlinear main effects. In this work we address this problem by proposing a Cross-validated Ensemble of Kernels (CVEK) that learns the space of appropriate functions for the main effects using a cross-validated ensemble approach. With a carefully chosen library of base kernels, CVEK flexibly estimates the form of the main-effect functions from the data, and encourages test power by guarding against over-fitting under the alternative. The method is motivated by a study on the interaction between metal exposures \textit{in utero} and maternal nutrition on children's neurodevelopment in rural Bangladesh. The proposed tests identified evidence of an interaction between minerals and vitamins intake and arsenic and manganese exposures.
Results suggest that the detrimental effects of these metals are most pronounced at low intake levels of the nutrients, suggesting nutritional interventions in pregnant women could mitigate the adverse impacts of \textit{in utero} metal exposures on children’s neurodevelopment.
\end{abstract}
}
\noindent%

\vfill
\thispagestyle{empty}
\newpage
\setcounter{page}{1}

\section{Introduction}

Investigation of the interplay between multiple lifestyle, biological, and environmental factors contributing to disease risk is a major goal in public health. Classic gene-environment and nutrition-environment studies focus primarily on the interaction between discrete factors \cite{manuck_gene-environment_2014, cornelis_gene-environment_2012}, or  between discrete factors and the linear effect of a few continuous measurements (e.g. \cite{lin_test_2016}). In recent years, however, recognizing the fact that populations are exposed to combinations of continuously-measured chemical and non-chemical factors that potentially have a nonlinear effect on outcome, there has been increasing interest in the best ways to statistically quantify the complex interplay of these continuous, nonlinear effects on health.   

In this work we analyze data from a birth cohort study on the interaction between \textit{in utero} exposure to a metal mixture and maternal nutrition intake during pregnancy on children's neurodevelopment in rural Bangladesh \cite{gleason_contaminated_2014, kile_prospective_2014}. Bangladesh has been  experiencing unparalleled levels of arsenic (As) other toxic metal poisoning through contaminated groundwater \cite{raessler_arsenic_2018}. Bangladesh also has rates of undernutrition that are among the highest in the world \cite{unicef_2016_2016}.
A recent study \cite{valeri_joint_2017} assessed the relationships between the arsenic (As), manganese (Mn), and lead (Pb) metal pollution mixture and infant neurodevelopment in Bangladesh, and has detected nonlinear, inverted-U shaped exposure–response relationships that differ among population subgroups, suggesting a role of additional cultural/behavioral factors, such as dietary habit, in affecting the impact of this metal mixture on children's health.  One possible factor impacting these environmental effects is maternal nutrition during pregnancy.  At vulnerable stages of fetal development, mother's overall nutrition intake may exacerbate adverse effects of chemical stressors.  Specific nutrients may modify chemical effects because of their influence on the metabolism of the chemicals, on epigenetic programming in response to the chemicals, or through other mechanisms that vary by metal or by outcome. 
To answer this question,  a companion nutritional study \cite{lin_validation_2017} was conducted to collect data on mother's nutrition intake during pregnancy, measuring the level of nutrition intake of 27 nutrients grouped in five nutrition categories (macronutrient, minerals, (pro-)vitamin As, vitamin Bs, and other vitamins), thereby providing an unique opportunity for researchers to quantitatively investigate the effect modification between  nutrition intake during pregnancy and \textit{in utero} metal exposures on infant development. 

The Bangladesh study posed two challenges that are common in many modern data science applications: (1) high dimensionality of the interaction, as the interaction term contains second- and higher-order interactions between 27 nutrients and 3 metal exposures, and (2) the nonlinearity of the underlying exposure-response relationship, whose  mathematical properties are unknown \textit{a priori}. In such a scenario, linear-model based methods are known to suffer from misspecification of the main effects model for nutrients (that include nutrient-nutrient interactions) and metals (including metal-metal interactions) even under the null of no nutrient-metal interactions, leading to inflated Type I error and reduced test power \cite{tchetgen_tchetgen_robustness_2011, cornelis_gene-environment_2012, wu_kernel_2013}. To boost efficiency and incorporate nonlinearities in the exposure-response relationship, a recent line of research has focused on constructing interaction tests based on kernel machine regression (KMR) \cite{scholkopf_learning_2002, rasmussen_gaussian_2006}. Building on the success of  previous kernel testing literature \cite{liu_semiparametric_2007, wu_rare-variant_2011}, these tests model the main-effect and interaction-effect functions as elements in reproducing kernel Hilbert spaces (RKHS) generated by pre-specified kernel functions, and build the hypothesis test by  re-parametrizing the kernel machine regression as a linear mixed model \cite{liu_semiparametric_2007}. In this framework,  the interaction term is an additional random effect term controlled by an univariate garrote parameter, on which one can construct a variance-component score test \cite{lin_variance_1997} for a test of the null hypothesis of no interaction. 
Successful applications of such tests include targeted gene effect identification in genetic pathway analysis \cite{maity_powerful_2011}, gene–gene interaction detection in genome-wide association study \cite{li_gene-centric_2012}, and also in gene-environment interaction studies with discrete factor such as gender \cite{broadaway_kernel_2015} and risk indicators of cardiovascular disease  \cite{ge_kernel_2015}. 

Applications of interaction tests involving sets of multiple continuous measurements with nonlinear effects, however, remain rare.
The key challenge impeding the success of interaction tests in continuous settings lies in  designing a proper kernel function for the multi-dimensional, nonlinear main-effect functions of unknown form. 
The kernel functions for the main effect terms need to generate a RKHS that is rich enough to contain the  main-effect functions under the null, 
while at the same time be sufficiently structured to maintain power for detecting interactions. 
Earlier work \cite{liu_semiparametric_2007, maity_powerful_2011} approached this problem by selecting the kernel from an assumed  parametric family (e.g. the Gaussian radial basis functions (RBF)) through maximum likelihood estimation, risking the specification of overly strong assumptions for these nonlinear functions. 
More recent approaches alleviate assumptions on the data-generation mechanism by incorporating multiple candidate kernels into the analysis, treating the kernel function as a weighted combination of candidate kernels, and learning kernel weights by maximizing various objective functions such as centered kernel alignment \cite{cortes_algorithms_2012} or by an $L_1$-regularized model likelihood \cite{seoane_pathway-based_2014}. However, designed primarily to maximize predictive accuracy, such procedures can be overly flexible under the alternative and potentially result in hypothesis tests with low power \cite{zhao_testing_2015}. 
Permutation tests are another popular approach for alleviating the issue of kernel misspecification \cite{cai_kernel_2011, zhao_testing_2015}; however, constructing a permutation procedure for an interaction test is usually not possible in observational studies, since the gene-environment independence condition tends to  not hold \cite{buzkova_permutation_2011}.

In this article, we propose a new approach to test for the interaction effect between groups of continuous features, each having potentially complex main effect functions relating outcome to that set of exposures. Built under the framework of kernel machine regression, we address the issue of kernel misspecification by deploying an ensemble of candidate kernels, and carefully design the ensemble strategy so that it minimizes the generalization error of the overall ensemble \cite{elisseeff_leave-one-out_2002}. Consequently, the proposed test automatically estimates the form of the kernel under the null from the data and guards against overfitting the interaction effect under the alternative, resulting in a powerful test that is robust under a wide range of data generation mechanisms. 
As we discuss in Section \ref{sec:cvek}, such a strategy results in an estimator that enjoys oracle property for ensemble selection and good generalization performance in limited samples, thereby achieving a powerful null-model estimator especially suitable for hypothesis testing in epidemiology studies. We term our method the Cross-Validated Ensemble of Kernels (CVEK). 
In Section \ref{sec:simu},  we illustrate the robustness of our method by conducting simulation studies that evaluate the finite-sample performance (Type-I error and power) under a range of data-generating scenarios and compare the performance of the proposed approach with other popular interaction tests. Finally, in Section \ref{sec:data_anal}, we apply our method to data from the Bangladesh reproductive cohort study  \cite{gleason_contaminated_2014, kile_prospective_2014} to investigate the interaction between  mother's daily nutrient intake and \textit{in-utero} exposure to an environmental metal mixture (As, Mn and Pb) on children’s neurodevelopment. 

\section{Background on the Study of Nutrition-Metal Interactions on Neurodevelopment in Bangladesh Children}
\label{sec:study}

The Bangladesh Reproductive Cohort Study (Project Jeebon) was initiated in 2008 to investigate the effects of prenatal and early childhood exposure to As, Mn and Pb on early childhood development. During 2008-2011, pregnant female participants (with gestational age < 16 weeks) were recruited from two rural health clinics operated by the Dhaka Community Hospital Trust (DCH) in the Sirajdikhan and Pabna Sadar upazilas of Bangladesh. During 2008-2013, data were collected at five time points spanning the entire perinatal and early childhood period, including: initial clinic visit (gestational age < 16 weeks, Visit 1); pre-delivery clinic visit (gestational age = 28 weeks, Visit 2), time of delivery (Visit 3), post-delivery clinic visit (infant age less than 1 month, Visit 4), and a post-natal follow-up visit (infant age between 20-40 weeks, Visit 5). 
Our central hypothesis is  that  
children born from mother who had lower nutrient intake will be the most susceptible to adverse effects of metal exposures.
 
Detailed procedures for data collection and measurement protocols have been documented previously \cite{gleason_contaminated_2014, kile_prospective_2014, valeri_joint_2017}. Briefly, background information on parent’s demographic status, including age, education, smoking history and socioeconomic status were collected through structured questionnaires at the two clinic visits during pregnancy (Visits 1-2). Information on infant’s biometric measurements, including sex, birth weight, length, head circumference, birth order and gestational age, were recorded at birth. Information on maternal medical history, maternal depression status (in Edinburgh Depression scale), maternal IQ (assessed using the Raven’s Progressive Indices \cite{raven_manual_1998}) were measured during the pregnancy visits (Visits 1-2), and an infant’s early childhood development, medical history, and quality of home environment (in terms of emotional, social, and cognitive stimulation, measured by HOME instrument score \cite{black_iron_2004} were measured during the follow-up visits (Visits 4-5), respectively.

Each infant’s exposure to multiple metals As, Mn and Pb (concentrations in $\mu g/dL$) during pregnancy were measured using blood samples from infant’s umbilical cord venous blood collected at the time of the birth. Mother’s overall nutrition intake status during pregnancy were measured for 27 nutrients derived from semi-quantitative Food Frequency Questionnaires (FFQs) specially adapted to Bangladeshi diet \cite{lin_associations_2017} at both the pre- and post-delivery visits (Visit 2 and 4). This instrument derives data on these 27 nutrients from measures of the consumption frequency (amount per week) of 42 food items during the 12-month period preceding delivery.   The nutrients measured can be grouped into 5 categories including macro-nutrients (5 nutrients: protein, fat, carbohydrate, dietary fiber and ash),  minerals (8 nutrients: calcium, iron, magnesium, phosphorus, potassium, sodium, zinc and copper), vitamin A and provitamin As (6 nutrients: vitamin A, retinol, beta-carotene equivalents, alpha-carotene, beta-carotene, and cryptoxanthin), vitamin B (5 nutrients: thiamin (B1), riboflavin (B2), niacin (B3), vitamin B6 and folate (B9)), and other vitamins (3 nutrients: vitamin C (i.e. L-ascorbic acid), vitamin D, and vitamin E). Finally, infant’s neurodevelopmental outcomes were assessed at 20–40 months of age (Visit 5) using a translated and culturally-adapted version of the Bayley Scales of Infant and Toddler Development, Third Edition (BSID-III) including five cognitive domains: cognitive, receptive language, expressive language, fine motor and gross motor. 

\section{Model and Inference}
\label{sec:review}

Assume we observe data from $n$ independent subjects. For the $i^{th}$ subject, let $y_i$ be a continuous response, $\bx_i$ be the set of $p$ baseline covariates that can be entered into the model linearly, and $\bz_i$ be the set of $q$ continuous covariates that have a nonlinear effect on $y_i$. Furthermore, we assume that there exists a grouping structure among the $\bz_i$ covariates such that $\bz_i = \{ \bz_{1,i}, \bz_{2,i} \}$, where the $m^{th}$ group $\bz_{m,i} \in \real^{q_m}$ contains $q_{m}$ covariates, $m=1,2$. We discuss the generalization to the case of more than two groups in $\bz_i$ in Section \ref{sec:test_nuis}. 

We assume that the outcome $y_i$ depends on covariates $\bx_i$, $\bz_i$  through the model:
\begin{align}
\label{eq:main}
y_i = \bx_i^T\bbeta + h(\bz_i) + \epsilon_i  \qquad
\mbox{where } \epsilon_i \stackrel{iid}{\sim} N(0, \sigma^2),
\end{align}
where $\bbeta$ is a $p \times 1$ vector of unknown coefficients for background covariates,  $h(\bz_i): \real^{q} \rightarrow \real$ is an unknown continuous function describing the effect of $\bz_i$, and $\epsilon_i$ is random noise that is independently and identically distributed as $N(0, \sigma^2)$. For identifiability purpose, $h$ is assumed to be square-integrable and subject to the constraint $\int_{\real^q} h(\bz) d\bz =0$. 

Our main objective in this work is to test for the interaction between two chosen sets of covariates in $\bz_i = \{ \bz_{1,i}, \bz_{2,i} \}$, while accounting for interactions within each covariate set. Without loss of generality, consider testing for the interaction between $\bz_{1,i}$ and $\bz_{2,i}$. Then our hypothesis is:
\begin{align}
\label{eq:null}
H_0: \quad  & h \in  \Hsc_{12}^\perp,
\end{align}
where $\Hsc_{12}$ is the space of "pure interaction" functions that contain only the interaction effect between $(\bz_{1, i}, \bz_{2,i})$.  That is, under the null hypothesis, $h(\bz)$ may depend on the individual main effects of $\bz_{1,i}$, $\bz_{2,i}$, but does not depend on the interaction effect of the set pair $(\bz_{1,i}, \bz_{2,i})$. 

We take the penalized likelihood approach to estimate parameters $(\bbeta, h)$. Namely, we first specify $\Hsc$ the candidate space and $\lambda$ the penalty parameter, then estimate parameters $\hat{\btheta} = (\hat{\bbeta}, \hat{h})$ by minimizing the penalized negative log likelihood:
\begin{align}
\label{eq:erm}
(\hat{\bbeta}, \hat{h}) = \underset{\bbeta \in \real, \; h \in \Hsc}{argmin}  \; L_{\lambda}(\bbeta, h), \qquad \mbox{where } 
\quad
L_{\lambda}(\bbeta, h) = 
\sum_{i=1}^n ||y_i - \bx_i \bbeta + h(z_i)||^2 + \lambda ||h||^2_{\Hsc}.
\end{align}

We model $\Hsc$ using Kernel Machine Regression (KMR) \cite{scholkopf_learning_2002}. Specifically, we assume $\Hsc$ to be a Reproducing Kernel Hilbert Space (RKHS) generated by a positive-definite kernel function $k(\bz_i, \bz_i')$, such that any $h \in \Hsc$ can be expressed in terms the kernel function as $f(\bz_i) = \inner{f, k(\bz_i, .)}_\Hsc$. Then by the Representer theorem \cite{burges_advances_1999}, if we define $\by_{n \times 1} = [y_1, \dots, y_n]^T$, $\bX_{n \times p} = [\bx_1^T, \dots, \bx_n^T]^T$, $\balpha =  [\alpha_1, \dots, \alpha_n]^T$ and also denote $\bK_{n \times n}$ the kernel matrix with its $(i, j)^{th}$ element to be $\bK_{i, j} = k(\bz_i, \bz_j)$, then (\ref{eq:erm}) can be re-written as
\begin{align}
\label{eq:lskm}
(\hat{\bbeta}, \hat{\balpha}) = \underset{\bbeta \in \real, \; \balpha \in \real^n}{argmin}  \; L_{\lambda}(\bbeta, \balpha), \qquad \mbox{where }
L_{\lambda}(\bbeta, \balpha) = 
||\by - \bX\bbeta - \bK\balpha||^2 + \lambda \; \balpha^T \bK \balpha
\end{align}
Furthermore, if we define $\tau = \frac{\sigma^2}{\lambda}$, $\bh^*$ can arise exactly from a linear mixed model (LMM) \cite{liu_semiparametric_2007}
\begin{align}
\label{eq:lmm}
\by = \bmu + \bh + \bepsilon 
\qquad \mbox{where} \qquad
\bh \sim N(\bzero, \tau \bK) \qquad 
\bepsilon \sim N(\bzero, \sigma^2 \bI).
\end{align}

\subsection{A Variance Component Test for Kernel Interaction}
\label{sec:test_int}

Under the LMM formulation of Kernel Machine regression in (\ref{eq:lmm}), Maity and Lin (\cite{maity_powerful_2011}) built a general test for the hypothesis $H_0: h \in \Hsc_0$ by assuming that $h$ lies in a RKHS generated by a \textit{garrote kernel function} $k_\delta(\bz, \bz')$, which is constructed by attaching an extra \textit{garrote parameter} $\delta$ to a regular kernel function. When $\delta = 0$, the garrote kernel function $k_0(\bz, \bz') = k_\delta(\bz, \bz') \Big|_{\delta = 0}$ generates exactly $\Hsc_0$ the space of functions under the null hypothesis. The authors further proposed a REML-based variance component score test for $H_0$.
In order to adapt the above approach to the hypothesis for interaction $H_0: \; h \in  \Hsc_{12}^\perp$, we construct the garrote kernel function $k_\delta(\bz, \bz')$ by building its corresponding RKHS for the main-effect and interaction space using the tensor-product construction \cite{gu_smoothing_2013}. Briefly, let $\bone = \{f | f \propto 1 \}$ be the RKHS of constant functions with kernel function $k(\bz, \bz')=1$, and let $\Hsc_m$ be the RKHS of centered functions (i.e. $\int f(\bz_{m}) \, d\bz_{m} = 0$) with domain on the  $m^{th}$ covariate set $\bz_{m}$, then $\Hsc$ adopts below orthogonal decomposition:
\begin{align*}
\Hsc &= 
\bone \oplus \Big\{ \Hsc_1 \oplus \Hsc_2 \Big\}
\oplus 
\Big\{ \Hsc_1 \otimes \Hsc_2 \Big\} 
\end{align*}
where $\Hsc_0 = \Hsc_1 \oplus \Hsc_2$ is the space of main-effect functions that does not contain the $(\bz_{1,i}, \bz_{2,i})$ interaction, and $\Hsc_{12}^\perp = \Hsc_1 \otimes \Hsc_2$ is ``pure interaction" space whose elements contain only the interaction effect between $(\bz_{1,i}, \bz_{2,i})$. Correspondingly, the garrote kernel function for interaction can be constructed as
\begin{align}
\label{eq:k_g}
k_\delta(\bz, \bz') &= k_0(\bz, \bz') + \delta * k_{12}(\bz, \bz')
\end{align}
where $k_0(\bz, \bz') = k_1(\bz, \bz') + k_2(\bz, \bz')$ and $k_{12}(\bz, \bz') = k_1(\bz, \bz') * k_2(\bz, \bz')$.

Under the above form of the garrote kernel function, the element of the null derivative kernel matrix $\bK_0$ is $\deriv{\delta} k_\delta(\bz, \bz') = k_{12}(\bz, \bz')$, i.e. the null derivative kernel matrix $\partial \bK_{0}$ is simply the kernel matrix  $\bK_{12}$ that corresponds to the interaction space. Therefore the test statistic is: 
\begin{align}
\label{eq:testat}
\hat{T}_0 &= \hat{\tau} * (\by - \bX\hat{\bbeta})^T \bV_0^{-1} \;  \bK_{12} \; \bV_0^{-1} (\by - \bX\hat{\bbeta}).
\end{align}

The null distribution of $\hat{T}_0$ is a mixture of chi-squares that can be approximated using a scaled chi-square distribution $\kappa \chi^2_{\nu}$ using either Satterthwaite-Welch method \cite{zhang_hypothesis_2003} or other higher-moment approximations \cite{bodenham_comparison_2016}.

\subsection{Generalization to Multiple Groups with Nuisance Interaction}
\label{sec:test_nuis}

Our description so far assumes there exists no nuisance interaction  terms in the model $y = \bx^T\bbeta + h(\bz) + \epsilon$. However, in more realistic scenario, $\bz$ usually exhibits complex hierarchical structure subsuming multiple groups, and it is often of interest to test only for the interaction between two small subgroups of $\bz$, leaving other interactions as nuisance effect to be accounted for by the null model. For example, consider the case of nutrition-environment interaction in Bangladesh birth cohort, $\bz_i$  is the $30 \times 1$ vector of during-pregnancy exposure to 27 nutrients and 3 metal pollutants, corresponding the grouping structure $\bz_i = \{\bz_{\tt metal}, \bz_{\tt nutr}\}$, where  $\bz_{\tt nutr}$ is further divided into $\bz_{\tt nutr} = \{\bz_{\tt macro}, \bz_{\tt mineral}, \bz_{\tt vitA}, \bz_{\tt vitB}, \bz_{\tt vitO}\}$. Therefore, when testing for the interaction between metal mixture exposures and a specific nutrient group of interest, care should be given to formulate $h(\bz_i)$ such that it not only explicitly characterizes the interaction of interest, but also account for all nuisance interactions among other $\bz_i$ subgroups. 

More specifically, assume $\bz_i = \{\bz_1, \bz_2, \bz_3\}$, when testing for the interaction between $\bz_1$ and $\bz_2$, the \textit{nuisance interactions} between $\bz_1$ and $\bz_3$, as well as between $\bz_2$ and $\bz_3$, should also be included in the null model. To this end, following the tensor-product construction shown in Section \ref{sec:test_int}, we adopt the following orthogonal decomposition of $h(\bz)$:
\begin{align*}
h(\bz_1, \bz_2, \bz_3) &= 
\bigg[
h_1(\bz_1) + h_2(\bz_2) + h_3(\bz_3)
\bigg] + \bigg[
h_{12}(\bz_1, \bz_2) + h_{13}(\bz_1, \bz_3) + h_{23}(\bz_2, \bz_3) 
\bigg] +
h_{123}(\bz_1, \bz_2, \bz_3)
\end{align*}
Under such construction, the null hypothesis of no interaction corresponds to $h_{12}$ and  $h_{123}$  equaling zero, i.e.
\begin{align*}
H_0: \qquad h &= h_1 + h_2 + h_3 + h_{13} + h_{23} \\
H_a: \qquad h &= h_1 + h_2 + h_3 + h_{13} + h_{23} + h_{12} + h_{123},
\end{align*}
and the corresponding garrote kernel for $h \in \Hsc$ is $k_\delta(\bz, \bz') = k_0(\bz, \bz') + \delta * k_a(\bz, \bz')$, where $k_{0} = k_1 + k_2 + k_3 + k_{13} + k_{23}$ and $k_{a} = k_{12} + k_{123}$. Here the $k_1$, $k_2$, and $k_3$ are the reproducing kernels for the main-effect space of $\bz_1$, $\bz_1$, $\bz_3$, respectively, and the higher-order interaction kernels are constructed as $\forall (i, j), k_{ij} = k_i * k_j$ and $k_{123} = k_1 * k_2 * k_3$ similar to Section \ref{sec:test_int}. Consequently, denoting $\bK_i$ as the kernel matrix corresponding to $k_i$, the null kernel matrix $\bK_0$ and the interaction kernel matrix $\bK_{12}$ are
\begin{alignat*}{2}
& \bK_0 && = \bK_1 + \bK_2 + \bK_3 + \bK_1 \circ \bK_2 + \bK_2 \circ \bK_3
\\
& \bK_{12} && =  \bK_1 \circ \bK_2 + \bK_1 \circ \bK_2 \circ \bK_3.
\end{alignat*}
and the test statistic can be constructed as in (\ref{eq:testat}).

\section{Robust Effect Estimation using Cross-validated Ensemble}
\label{sec:cvek}


We motivate the importance of robust null model estimation by considering the possible impact of a misspecified null kernel function $k_0$ on the performance of the resulting hypothesis test. Specifically, we express the test statistic $\hat{T}_0$ in (\ref{eq:testat}) in terms of the model residual $\hat{\bepsilon} = \by - \hat{\bmu} - \hat{\bh}$:
\begin{align}
\label{eq:score_norm}
\hat{T}_0 & \; \propto \;
\hat{\bepsilon}^T  \bK_{12} \hat{\bepsilon},
\end{align}
where we have used the fact $\bV_0^{-1}(\by - \hat{\bmu}) = (\hat{\sigma}^2)^{-1}(\hat{\bepsilon})$  \cite{harville_maximum_1977}. Therefore, the test statistic $\hat{T}_0$ is a scaled quadratic-form statistic that is a function of the model residual. If $k_0$ is too restrictive, model estimates will underfit the data under the null hypothesis, introducing extraneous correlation among the $\hat{\epsilon}_i$'s that yield inflated $\hat{T}_0$ values and deflated p-values under the null. Therefore, this approach will
yield an invalid test having inflated Type I error.  On the other hand, if $k_0$ is too flexible, model estimates will likely overfit the data in small samples, producing underestimated residuals, which leads to underestimated test statistics and overestimated p-values.
Accordingly, the resulting test will have low power.

The above observations motivate a kernel estimation strategy that is flexible in that it does not underfit under the null, yet stable so that it does not overfit under the alternative. To this end, we propose estimating $h$ using the convex ensemble of a library of fixed base kernels $\{k_d\}_{d=1}^D$:
\begin{align}
\label{eq:cvek}
\hat{h}(\bx) = \sum_{d=1}^D u_d \hat{h}_d(\bx) \qquad
\bu \in \Delta = \{\bu| \bu \geq 0, \bone^T\bu = 1 \},
\end{align}
where $\hat{h}_d$ is the kernel predictor generated by $d^{th}$ base kernel $k_d$. In order to maximize model stability, the ensemble weights $\bu$ are estimated to minimize the overall cross-validation error of $\hat{h}$. We term this method the \textit{Cross-Validated Kernel Ensemble} (CVKE). 
The exact algorithm proceeds in three stages as follows (see Algorithm \ref{alg:cvke} for summary): 

\textbf{Stage 1: Candidate Kernel Predictors}
For each basis kernel in the library $\{ k_d \}_{d=1}^D$, we first standardize the kernel matrix by its trace $\bK_d = \bK_d/tr(\bK_d)$, and then estimate the prediction based on each kernel as $\hat{\bh}_{d, \hat{\lambda}_d} = \bK_d (\bK_d + \hat{\lambda}_d \bI)^{-1} \by, d\in\{1, \dots, D\}$, where the tuning parameter $\hat{\lambda}_d$ is selected by minimizing the k$-$fold cross-validation error. We denote the estimate of the cross-validation error for $d^{th}$ kernel as $\hat{\epsilon}_d = \texttt{CV}\Big(\hat{\lambda}_d | \bK_d \Big)$. 

\textbf{Stage 2: Cross-validated Ensemble}
Using the estimated cross-validation errors $\{\hat{\epsilon}_d\}_{d=1}^D$, estimate the ensemble weights $\bu = \{u_d\}_{d=1}^D$ by minimizing the overall cross-validation error $\hat{\bepsilon}_{\bu} = \sum_{d=1}^D u_d \hat{\epsilon}_d$:
\begin{align*}
\hat{\bu} = \underset{\bu \in \Delta}{argmin} \; ||\sum_{d=1}^D u_d \hat{\epsilon}_d||^2, \qquad 
\mbox{where} \quad
\Delta = \{\bu| \bu \geq 0, \bone^T\bu = 1 \},
\end{align*}
and produce the final ensemble prediction $
\widehat{\bh}
= \sum_{d=1}^D \widehat{u}_d \widehat{\bh}_d 
= \sum_{d=1}^D \widehat{u}_d  \bA_{d, \hat{\lambda}_d} \by 
= \hat{\bA} \by$,
where $\hat{\bA} = \sum_{d=1}^D \hat{u}_d \bA_{d, \hat{\lambda}_d}$ is the ensemble hat matrix.

\textbf{Stage 3: Ensemble Kernel Matrix} 
Using the ensemble hat matrix $\hat{\bA}$, estimate the ensemble kernel matrix $\hat{\bK}$ by solving $\hat{\bK} (\hat{\bK} + \lambda \bI)^{-1} = \widehat{\bA}$. Specifically, if we denote $\bU_A$ and $\{\delta_{A,k}\}_{k=1}^n$ as the eigenvector and eigenvalues of $\widehat{\bA}$, respectively, then the ensemble kernel matrix $\widehat{\bK}$ adopts the form :
\begin{align}
\label{eq:K_ens}
\widehat{\bK} = \lambda_\bK * \bigg[
\bU_A diag \Big( \frac{\delta_{A, k}}{1 - \delta_{A,k}} \Big) \bU_A^T \bigg],
\end{align}
where we recommended setting $\lambda_\bK = min\Big(1,  \big( \sum_{k=1}^n \frac{\delta_{A, k}}{1 - \delta_{A,k}} \big)^{-1} \Big)$ (see Supplementary Section  \ref{sec:ekm_deriv}).

We remind readers that the CVEK's ensemble form (Step 2) belongs to the general class of model aggregation method known as \textit{convex aggregation} \cite{tsybakov_optimal_2003}, whose oracle property in model selection has been established both  asymptotically and in finite-sample \cite{van_der_laan_unified_2003, lecue_oracle_2012}. It can be also considered as a special case of \textit{ensemble of kernel predictors} (EKP) \cite{cortes_ensembles_2011}, whose generalization behavior is well characterized in terms of the rate of  eigenvalue decay of the base kernels. Consequently, under the null hypothesis, with a diverse set of base kernels, the CVEK ensemble converges in $O(\frac{1}{n})$ rate to the “oracle ensemble” made by an oracle that has access to infinite amount of validation data, thereby resulting in correct Type I error by mitigating null model misspecification. Under the alternative, by setting the diverse kernel library to be a mix of parametric kernels (linear, polynomial) and smooth kernels of exponential eigendecay rate (e.g. a collection of Gaussian RBF kernel with different fixed spatial smoothness parameters), CVEK converges to its asymptotic counterpart in $O(\frac{1}{n})$ rate if the data-generation function is indeed parametric, and in the "near-parametric" rate of $O(\frac{log(n)}{n})$ if the data-generation function is complex and nonlinear, thereby encouraging good test power by not overfitting the interaction effect due to fast generalization rate. The resulting ensemble kernel is therefore a strong candidate for a null model estimator that is suitable for hypothesis testing. We refer readers to Supplementary Section \ref{sec:theory} for detailed discussion.

\section{Numeric Studies}
\label{sec:simu}

We evaluate the finite-sample performance of the proposed interaction test in a simulation study that mimics a small-sample nutrition-environment interaction study.
We generate two groups of input features $(\bz_{i,1}, \bz_{i,2}) \in \real^{p_1} \times \real^{p_2}$ independently from a multivariate Gaussian distribution, reflecting each subject's level of exposure to $p_1$ environmental pollutants and the levels of a subject's intake of $p_2$ nutrients during the study. Throughout the simulation scenarios, we keep $n = 200$, and $p_1 = p_2 = 3$. We generate the outcome $y_i$ as:
\begin{align}
\label{eq:gen}
y_i &=  h_1(\bz_{i,1}) + h_2(\bz_{i,2}) + \delta * h_{12}(\bz_{i,1}, \bz_{i,2}) + \epsilon_i,
\end{align}
where $h_1, h_2, h_{12}$ are sampled from RKHSs $\Hsc_1, \Hsc_2$ and $\Hsc_1 \otimes \Hsc_2$, generated using a ground-truth kernel $k_{\texttt{true}}$. We standardize all sampled functions to have unit norm, so that $\delta$ represents the strength of interaction relative to the main effect. For each simulation scenario, we first generate data using $\delta$ and  $k_{\texttt{true}}$, and then use a   $k_{\texttt{model}}$ to estimate the null model and obtain p-value using the proposed test. We repeat each scenario 1000 times, and evaluated the test performance using the empirical probability $\widehat{P}(p \leq 0.05)$. 

In this study, we vary $k_{\texttt{true}}$ to produce data-generating functions $h_\delta(\bz_{i,1}, \bz_{i,2})$ with different smoothness and complexity properties, and varied $k_{\texttt{model}}$ to reflect different common modeling strategies for the null model in addition to using CVEK. We then evaluate how these two aspects impact the hypothesis test's Type I error and power. More specifically, we sample the data-generating function using $k_{\texttt{true}}$ from Mat\'{e}rn kernel family \cite{rasmussen_gaussian_2006}:
$$k(\br | \nu, \sigma) = 
\frac{2^{1 - \nu}}{\Gamma(\nu)}
\Big( \sqrt{2 \nu} \sigma ||\br||\Big)^\nu
K_{\nu} \Big( \sqrt{2 \nu} \sigma ||\br|| \Big), \qquad \mbox{where} \qquad \br = \bx - \bx',
$$
with two non-negative hyperparameters $(\nu, \sigma)$. For a function $h$ sampled using a Mat\'{e}rn kernel, $\nu$ determines the function's smoothness (i.e. degree of mean-square differentiability), and $\sigma$ determines the function's complexity in terms of spectral frequency \cite{rasmussen_gaussian_2006}.
In this work, we vary $\nu \in \{\frac{3}{2}, \frac{5}{2}, \infty\}$ to generate once-, twice, and infinitely-differentiable functions, and vary $\sigma \in \{0.5, 1, 1.5\}$ to generate functions with varying degree of complexity.

We consider 12 $k_{\tt model}$'s that are grouped into five model families (See Table \ref{tb:k_model} for a complete summary): (1)
\textbf{Polynomial Kernels} that is equivalent to polynomial ridge regression. In this work, we use the \textbf{linear} kernel $k_{\texttt{linear}}(\bx, \bx'|p) = \bx^T\bx'$ and \textbf{quadratic} kernel $k_{\texttt{quad}}(\bx, \bx'|p) = (1 + \bx^T\bx')^2$. (2) \textbf{Gaussian RBF Kernels}: $k_{\texttt{RBF}}(\bx, \bx'|\sigma) = exp(- ||\bx - \bx'||^2/\sigma^2)$ is a general-purpose kernel family that generates nonlinear, but very smooth (infinitely differentiable), functions. Under this kernel, we consider two hyperparameter selection strategies commonly seen in application: \textbf{RBF-Median} where we set $\sigma$ to the sample median of $\{||\bx_i - \bx_j||\}_{i \neq j}$, and \textbf{RBF-MLE}, which estimates $\sigma$ by maximizing the REML likelihood. (3) \textbf{Mat\'{e}rn} and (4) \textbf{Neural Network Kernels} are two flexible kernel families both containing a rich space of candidate functions. For Mat\'{e}rn kernel, we use \textbf{Matern 1/2}, \textbf{Matern 3/2} and \textbf{Matern 5/2}, corresponding to flexible models that is capable of approximating non-differentiable, once-differentiable, and twice-differentiable functions. Neural network kernels \cite{rasmussen_gaussian_2006}, on the other hand, represent a 1-layer Bayesian neural network with $\sigma$ being the prior variance on the hidden weights, and it is theoretically capable of approximate arbitrary continuous functions on the compact domain \cite{hornik_approximation_1991}. In this work, we let \textbf{NN 0.1}, \textbf{NN 1} and \textbf{NN 10} denote Bayesian networks with different prior constraints $\sigma \in \{0.1, 1, 10\}$. Finally, we evaluate the performance of the (5) \textbf{Cross-validated Kernel Ensemble} estimator we propose here. Specifically, we consider a CVEK estimator based on a Gaussian RBF kernel with $log(\sigma) \in \{-2, -1, 0, 1, 2\}$, which we label \textbf{CVEK-RBF}. Furthermore, to evaluate the consequence of  more flexible kernel families on ensemble behavior, we also consider \textbf{CVEK-NN}, which is a ensemble of neural network kernels with $\sigma \in \{0.1, 1, 10, 50\}$). 

The simulation results are presented graphically in Figure \ref{fig:res} and documented in detail in the Supplementary Section \ref{sec:simu_detail}. We first observe that for reasonably specified values of $k_{\texttt{model}}$, the proposed hypothesis test with CVEK estimator always has the correct Type I error and reasonable power. We also observe that the complexity of the data-generating function $h_\delta$ (\ref{eq:gen}) plays a role in test performance, in the sense that the power of the hypothesis tests increases as the Mat\'{e}rn $k_{\texttt{true}}$'s complex parameter $\sigma$ becomes larger, which corresponds to functions that put more weight on the simpler, slow-varying eigenfunctions in Bochner's spectral decomposition \cite{rasmussen_gaussian_2006}.

There exist clear differences in test performance between different model families. In general, polynomial models (\textbf{linear} and \textbf{quadratic}) appear to be too restrictive and underfit the data under both the null and the alternative, producing inflated Type I error and diminished power. On the other hand, lower-order Mat\'{e}rn kernels (\textbf{Mat\'{e}rn 1/2} and \textbf{Mat\'{e}rn 3/2}, dark blue lines) appear to be too flexible, due to their slow eigenvalue decay. Whenever data are generated from similarly or smoother $k_{\texttt{true}}$,  \textbf{Mat\'{e}rn 1/2} and \textbf{3/2} overfit the data and produce deflated Type I error and severely diminished power, even if the hyperparameter $\sigma$ is fixed at its true value. Comparatively, Gaussian RBF works well for a wider range of $k_{\texttt{true}}$'s, but only if the hyperparameter $\sigma$ is selected carefully. Specifically, \textbf{RBF-Median} (black dashed line) works generally well, despite being slightly conservative (i.e. lower power) when the data-generation function is smooth and of low complexity. \textbf{RBF-MLE} (black solid line), on the other hand, tends to underfit the data under the null and exhibits inflated Type I error, possibly because of the fact that $\sigma$ is not strongly identified when the sample size is modest \cite{wahba_spline_1990}. The situation becomes more severe as $h_\delta$ becomes rougher and more complex. In the more  extreme case of non-differentiable $h$ with $\sigma = 1.5$, the Type I error is inflated to as high as 0.24. Neural Network kernels also perform well for a wide range of $k_{\texttt{true}}$, and with the Type I error more robust to the specification of hyperparameters. 

Finally, the two ensemble estimators \textbf{CVEK-RBF} and \textbf{CVEK-NN} perform as well or better than the non-ensemble approaches for all $k_{\texttt{true}}$'s, despite being slightly conservative under the null.  As compared to \textbf{CVEK-NN}, \textbf{CVEK-RBF }appears to be slightly more powerful, validating recommendations in Section \ref{sec:cvek}.

\section{Data Analysis}
\label{sec:data_anal}

We use 
the proposed methods to test for nutrition-environment interactions in the Bangladesh birth cohort study (see Section \ref{sec:study} for complete study description). Our aim is to detect whether mother’s nutrient intake during pregnancy modifies the effect of metal mixture exposures on children’s early-stage fine motor BSID-III scores in the district of Pabna (n = 351). Specifically, our interest concentrates on the interaction between the mixture of As, Mn and Pb, and five major nutrient groups: \textit{macro-nutrient}, \textit{mineral}, \textit{vitamin A}, \textit{vitamin B} and the other vitamins (denoted as \textit{vitamin, other}). For each of the five nutrient groups, we test for the overall interaction between the selected group and the joint effect of the As, Pb, Mn mixture. We adjust for parent’s demographic status (age, education, smoking history), infant’s biometric measurements at birth (sex, birth weight, length, head circumference, birth order and gestational age), and quality of early-childhood home environment (HOME score, maternal depression scale, maternal IQ).

We compare our method with three existing approaches for testing high-dimensional interaction. The (1) \textit{Interaction Sequence Kernel Association Test} (iSKAT)\cite{lin_test_2016} is a baseline approach that assumes linear relationship between exposures and outcome. It estimates the null model using ridge linear regression and corresponds to the \textbf{linear} model in simulation. (2) The \textit{Gaussian Kernel Machine} test (GKM) \cite{maity_powerful_2011} estimates the null model using kernel machine regression with Gaussian RBF kernels and tunes the kernel hyperparameter by maximizing REML. It correspondes to the \textbf{RBF-MLE} model in simulation. Finally, the (3) GE-spline test \cite{he_set-based_2016} which uses the  generalized additive regression to model the nonlinear effect of environmental exposures using spline sieves. It can be considered as a special case of kernel machine regression with the kernel matrix constructed adaptively using spline basis functions \cite{kimeldorf_correspondence_1970}. In order to visualize the identified interaction  and thereby provide interpretable findings, we graphically summarize the multivariate interaction effects by examining the joint exposure-response surface between the principal components of the pollutant mixture and those for each nutrient group. 

\subsection{Nutrient - Mixture Interactions}
\label{sec:ana_all}  

Table \ref{tb:res_joint} presents p$-$values for the interaction between the overall metal mixture and each of the five nutrient groups. We conducted the proposed test using two types of CVEK models: an ensemble of seven neural network kernels with prior variance set between $log(\sigma^2) \in \{-3, -2, -1, 0, 1, 2, 3\}$ (denoted as CVEK-NN), and an ensemble of seven RBF kernels with bandwidth parameter set between $log(\sigma) \in \{-3, -2, -1, 0, 1, 2, 3\}$ (denoted as CVEK-RBF). We compared the results of each to those generated by iSKAT, GKM, and GE-Spline.  As shown in the table, most tests yielded strong evidence of interaction  ($p < 0.05$) for  \textit{vitamin A} and the \textit{vitamin, other} groups, as well as moderate evidence of interaction ($p <\approx 0.1$) for the \textit{mineral} and the \textit{vitamin B} groups. There was no evidence of an interaction between metal exposures and macro-nutrients.

Comparing the performance across different tests, we observed similar patterns for p-values  for CVEK-NN and CVEK-RBF, suggesting robustness in test performance with respect to the choice of the family of the base kernels. We also observed similar patterns for p-values between iSKAT (linear kernel) and GKM, the latter of which used a single RBF kernel with REML-based hyperparameter tuning. The results from these two tests are similar to those from  the CVEK tests for \textit{vitamin A} and \textit{vitamin, other} groups of nutrients, but are less powerful in detecting the interaction for the \textit{mineral} and the \textit{vitamin B} groups. This is consistent with our observation in Section \ref{sec:simu} that, when the true effect is nonlinear and  exhibits a moderate level of smoothness and low complexity  (a scenario that is likely to hold for the effect of environmental exposures, see Figure \ref{fig:res} (a)), the hypothesis test based on GKM performs similarly to that based on the iSKAT but is less powerful than the CVEK-based test. This reduction in power is possibly due to the overly strong smoothness assumption imposed by these two models. Finally, we notice that the performance of the test from the GE-spline model appears sub-optimal when compared to that of the other methods. GE-spline produced a much smaller p$-$value for the interaction for the A vitamins ($p=0.0084$), but much higher p$-$values for the other nutrient groups, failing to detect the interaction for the B vitamins and the other vitamins. We hypothesize that the observed instability of GE-spline is likely caused by the lack of fit of the null model, due to the difficulty in estimating multivariate splines in high dimensions.

\subsection{Visualization of Exposure-Response Surface}
\label{sec:dose-response}

\begin{figure}
\begin{subfigure}{.33\textwidth}
  \centering
  \includegraphics[width=\linewidth]{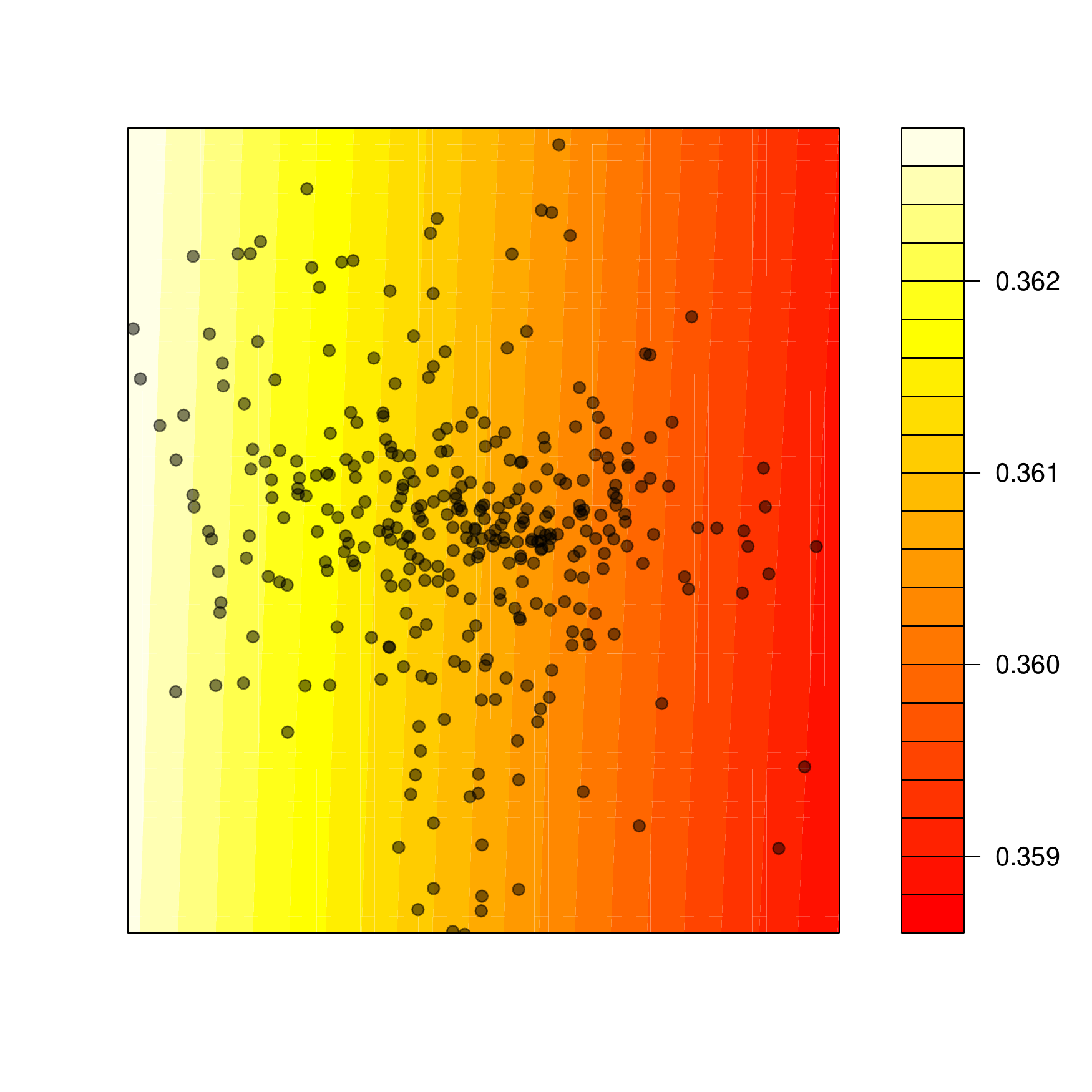}
  \caption{Mixture,  PC 1 vs {\tt macro} PC 1  }
\end{subfigure}
\begin{subfigure}{.33\textwidth}
  \centering
  \includegraphics[width=\linewidth]{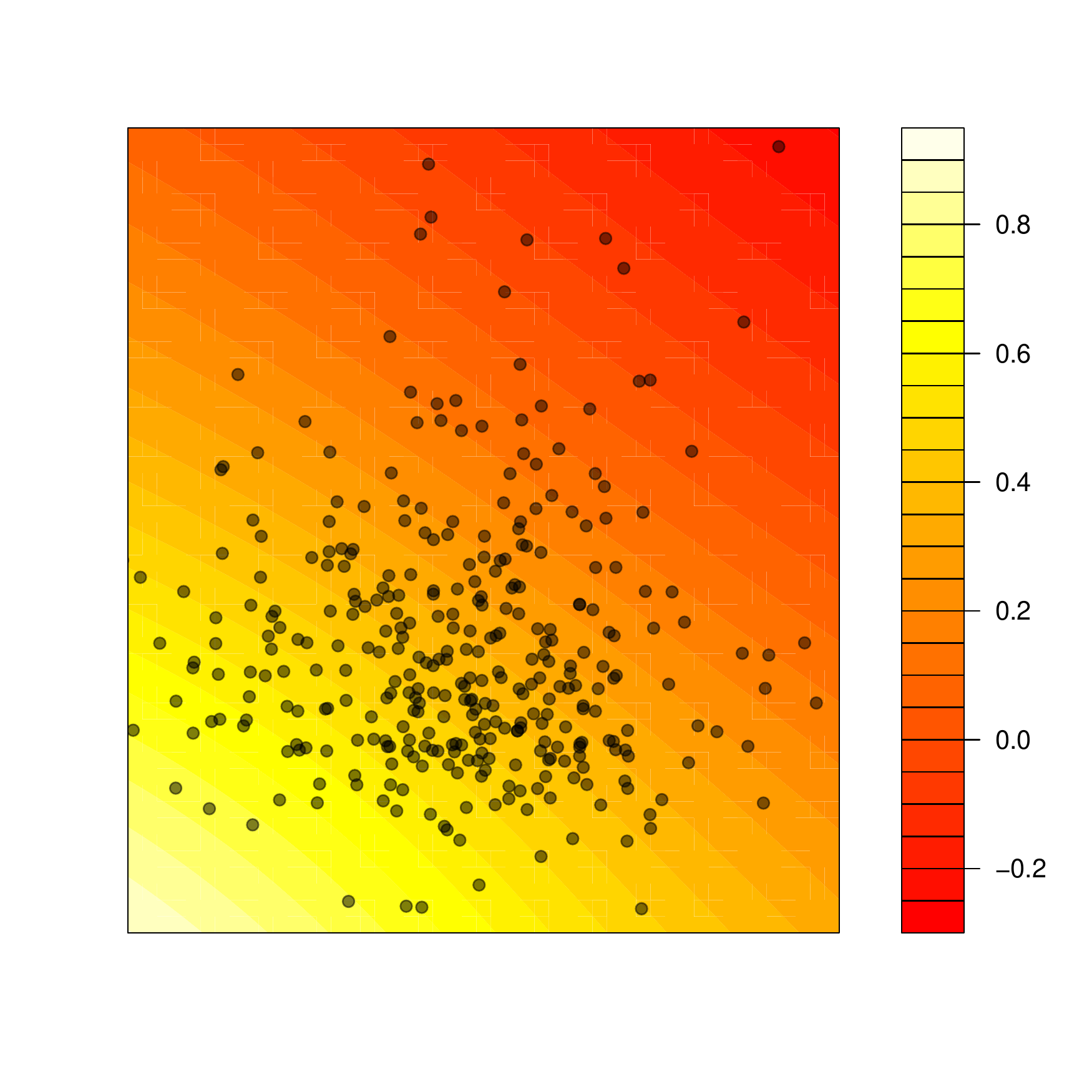}
  \caption{Mixture,  PC 1 vs {\tt macro} PC 2  }
\end{subfigure}
\begin{subfigure}{.33\textwidth}
  \centering
  \includegraphics[width=\linewidth]{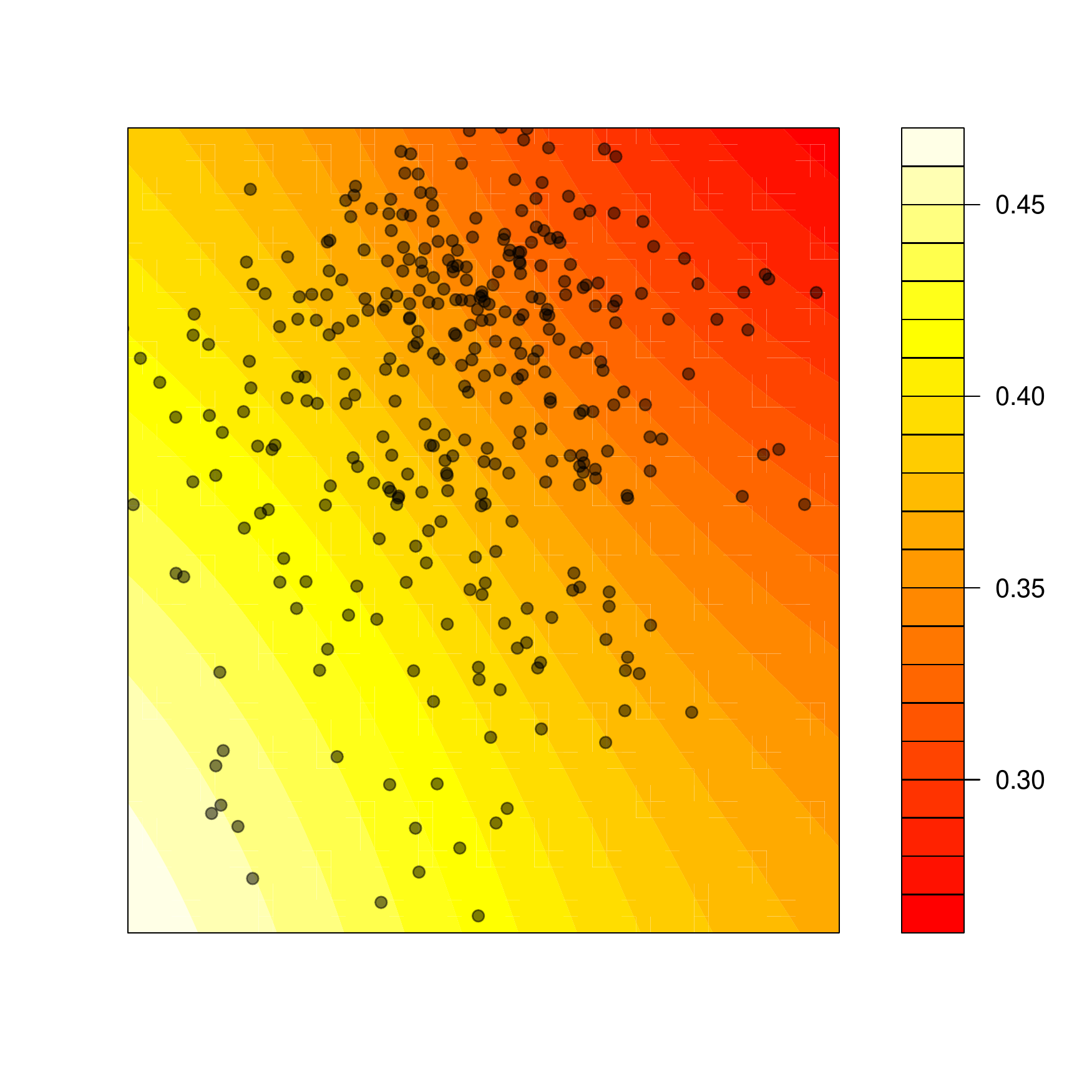}
  \caption{Mixture,  PC 1 vs {\tt macro} PC 3  }
\end{subfigure}

\begin{subfigure}{.33\textwidth}
  \centering
  \includegraphics[width=\linewidth]{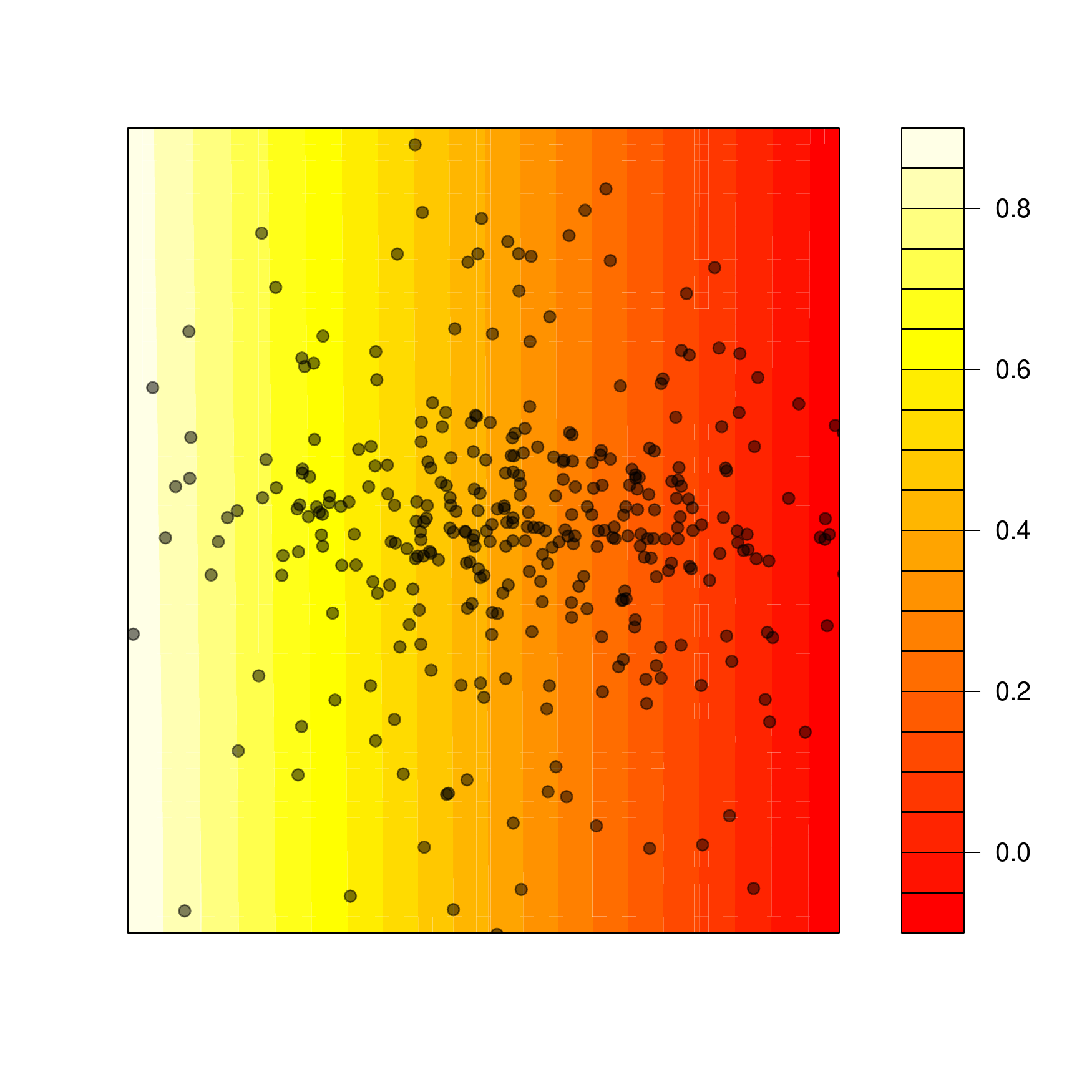}
  \caption{Mixture,  PC 2 vs {\tt macro} PC 1  }
\end{subfigure}
\begin{subfigure}{.33\textwidth}
  \centering
  \includegraphics[width=\linewidth]{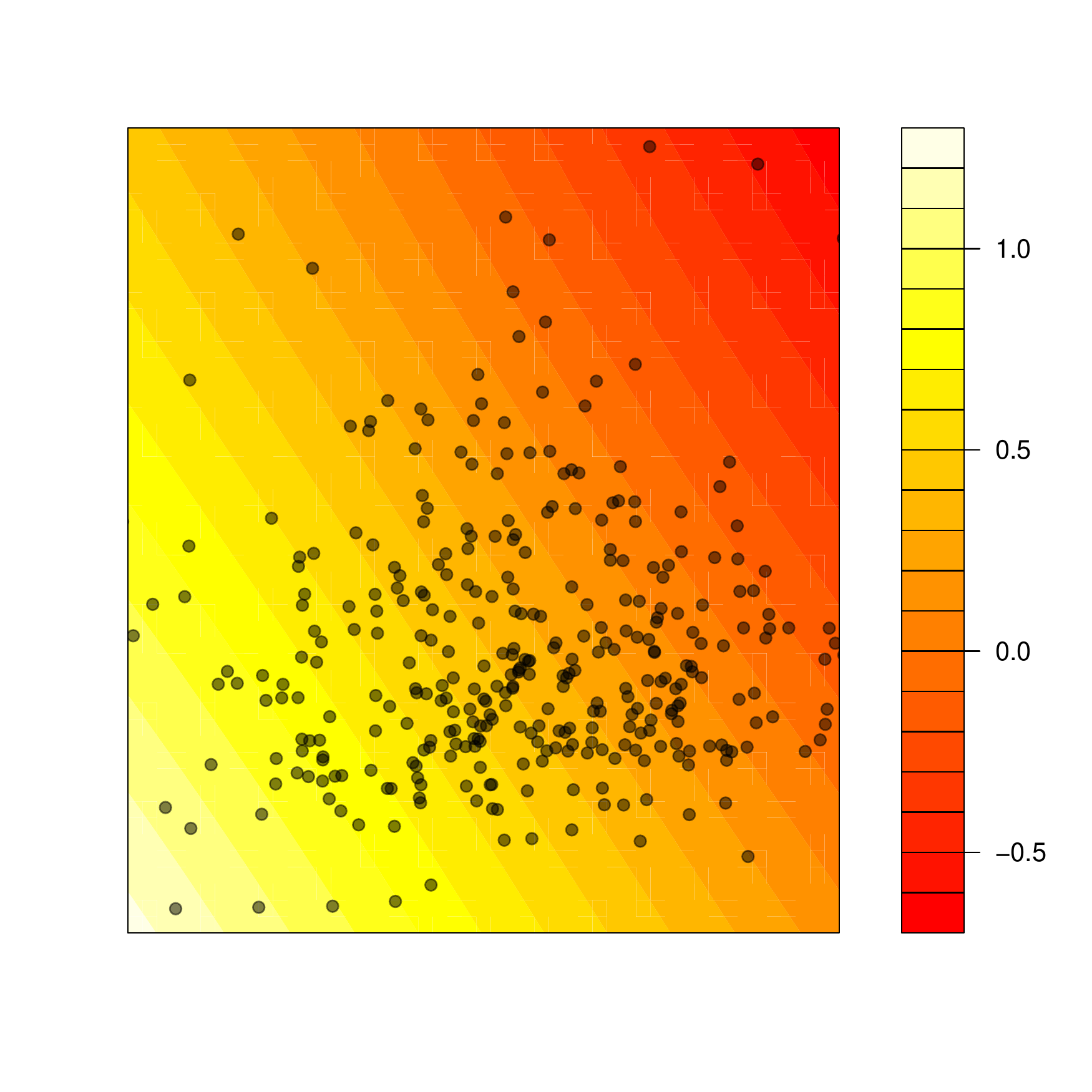}
  \caption{Mixture,  PC 2 vs {\tt macro} PC 2  }
\end{subfigure}
\begin{subfigure}{.33\textwidth}
  \centering
  \includegraphics[width=\linewidth]{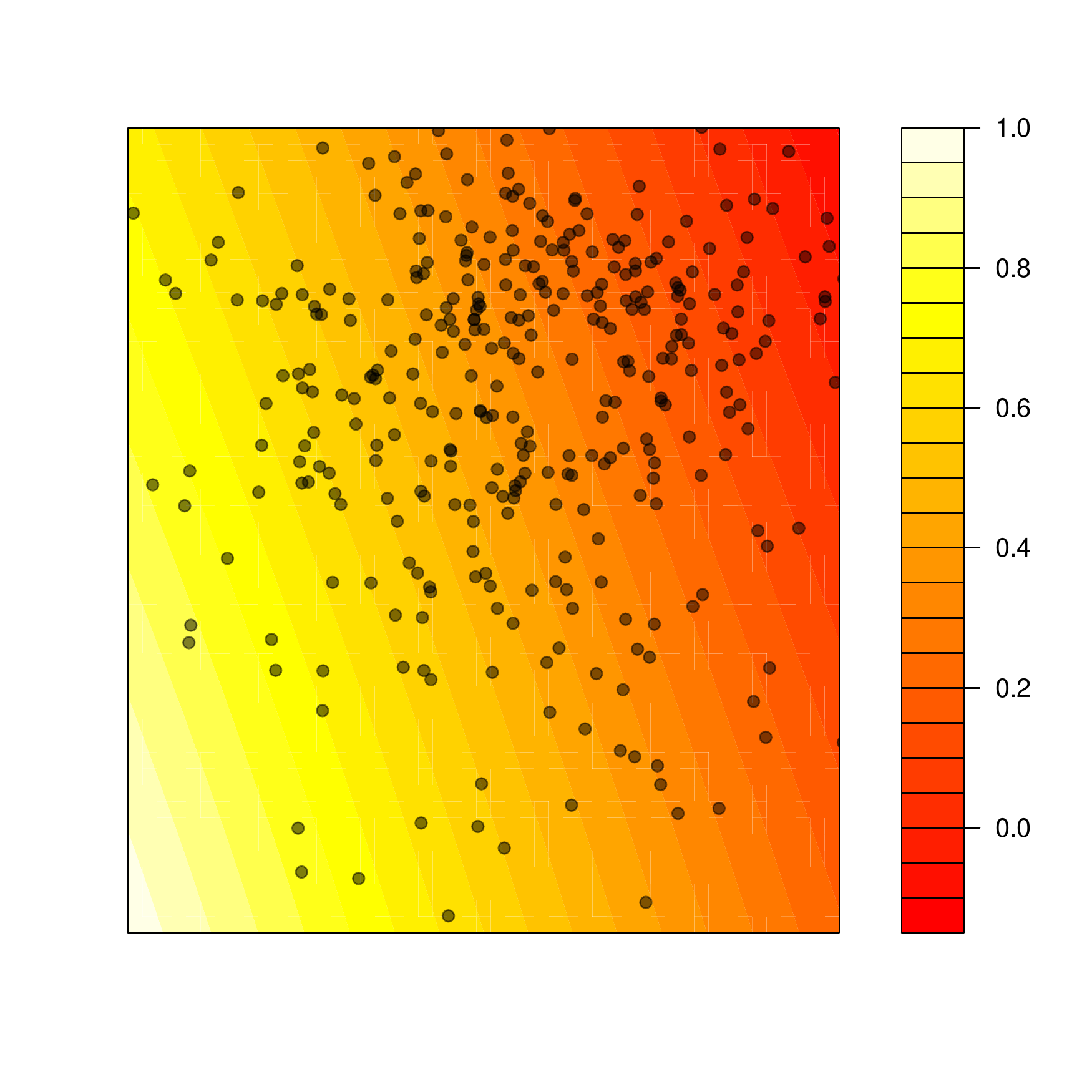}
  \caption{Mixture,  PC 2 vs {\tt macro} PC 3  }
\end{subfigure}

\begin{subfigure}{.33\textwidth}
  \centering
  \includegraphics[width=\linewidth]{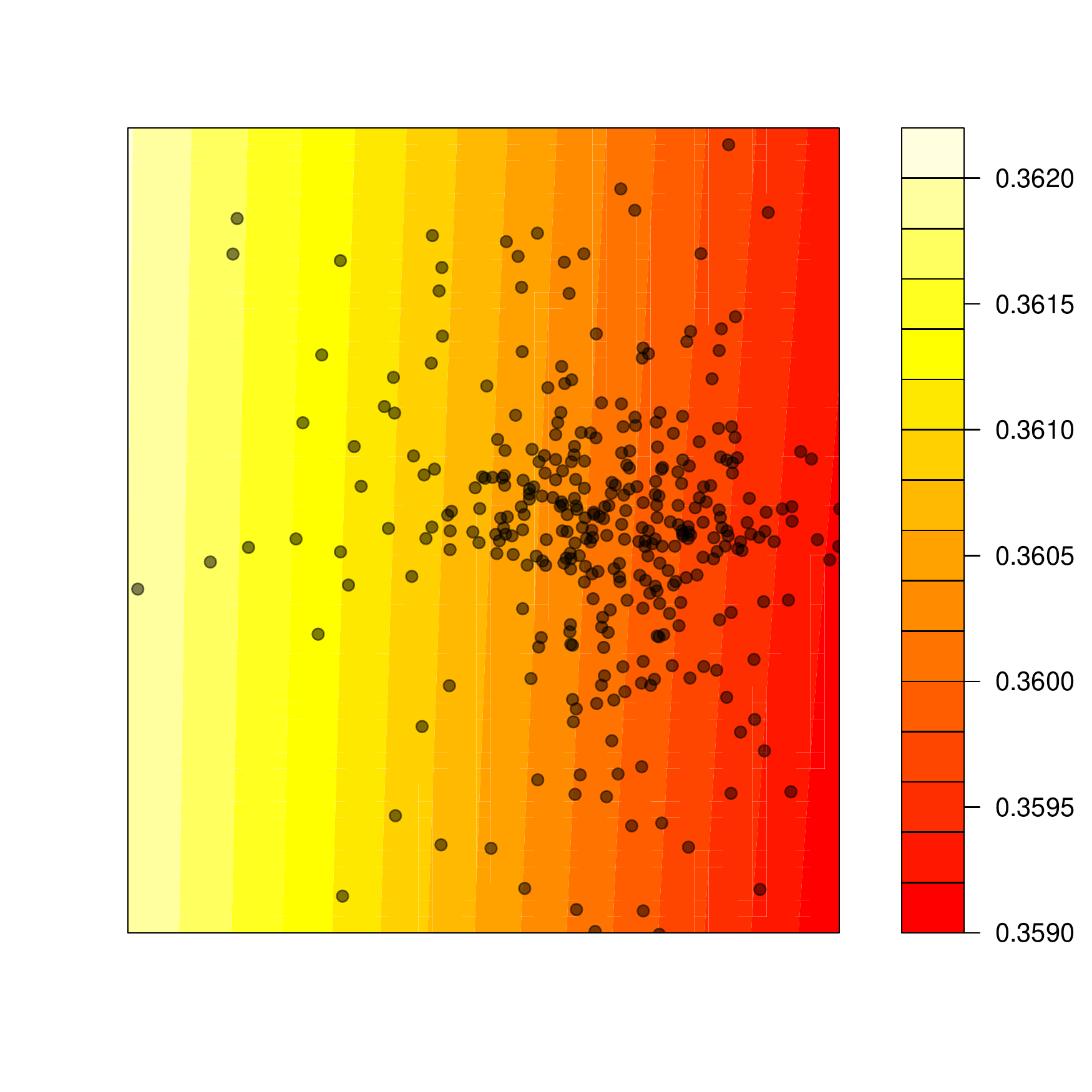}
  \caption{Mixture,  PC 3 vs {\tt macro} PC 1  }
\end{subfigure}
\begin{subfigure}{.33\textwidth}
  \centering
  \includegraphics[width=\linewidth]{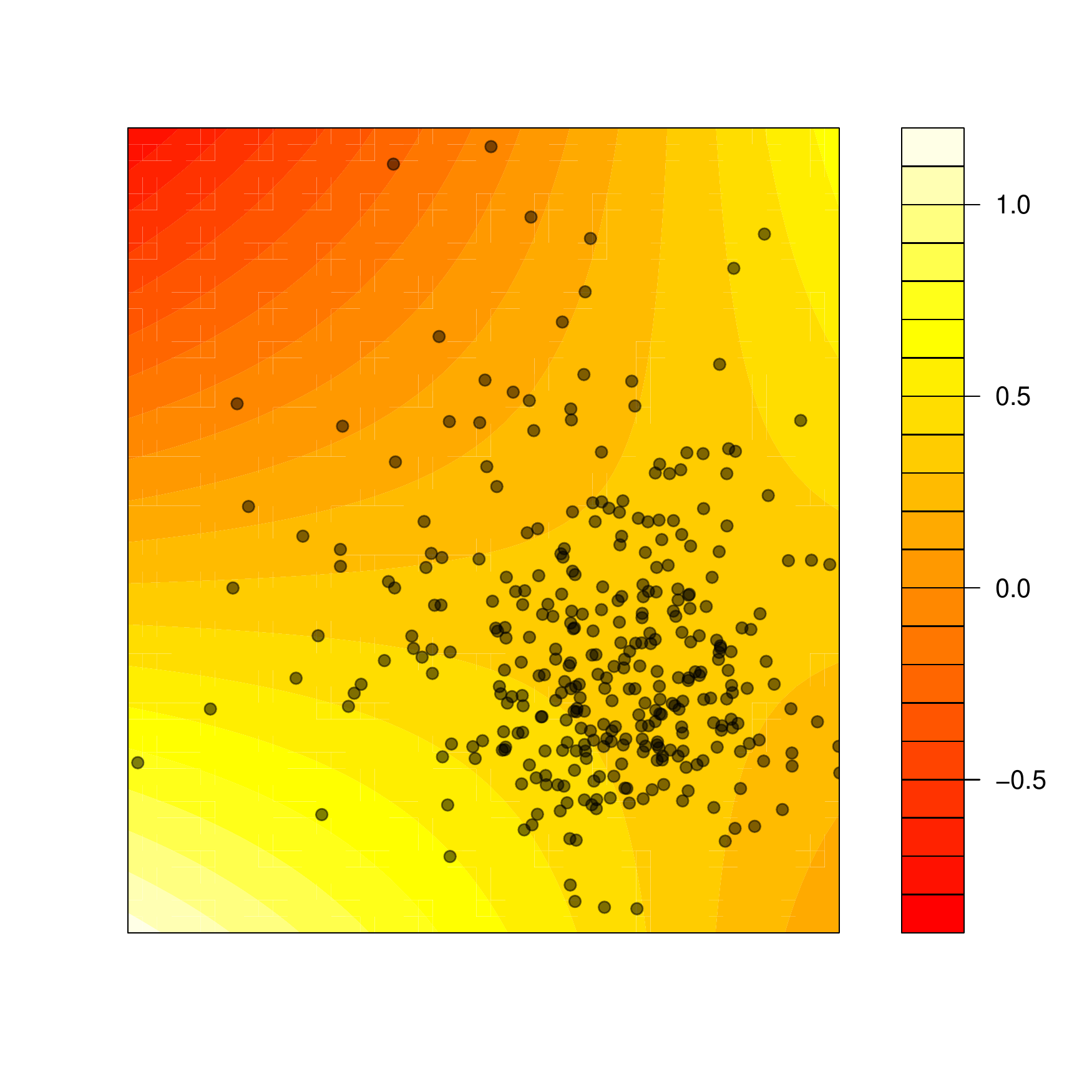}
  \caption{Mixture,  PC 3 vs {\tt macro} PC 2  }
\end{subfigure}
\begin{subfigure}{.33\textwidth}
  \centering
  \includegraphics[width=\linewidth]{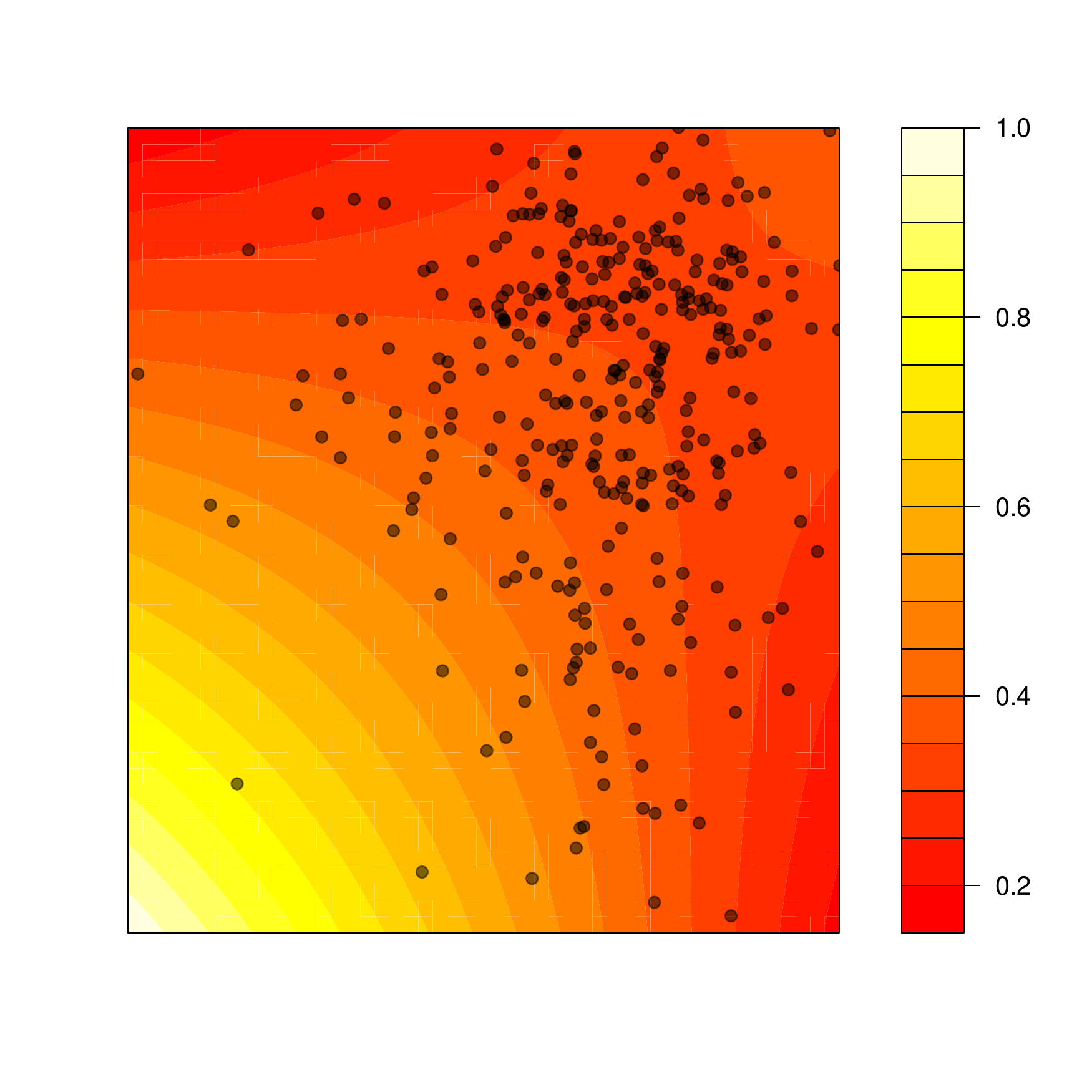}
  \caption{Mixture,  PC 3 vs {\tt macro} PC 3  }
\end{subfigure}

\caption{Interaction between joint mixture and macronutrient by principal components \\ 
The top 3 PCs for pollutants accounts for $42.60\%$, $37.34\%$, $20.05 \%$ of total variation,\\
The top 3 PCs for {\tt macro} accounts for $63.54 \%$, $28.46 \%$ and $7.36\%$ of total variation.}
\label{fig:int_macro}
\end{figure}

\begin{figure}
\begin{subfigure}{.33\textwidth}
  \centering
  \includegraphics[width=\linewidth]{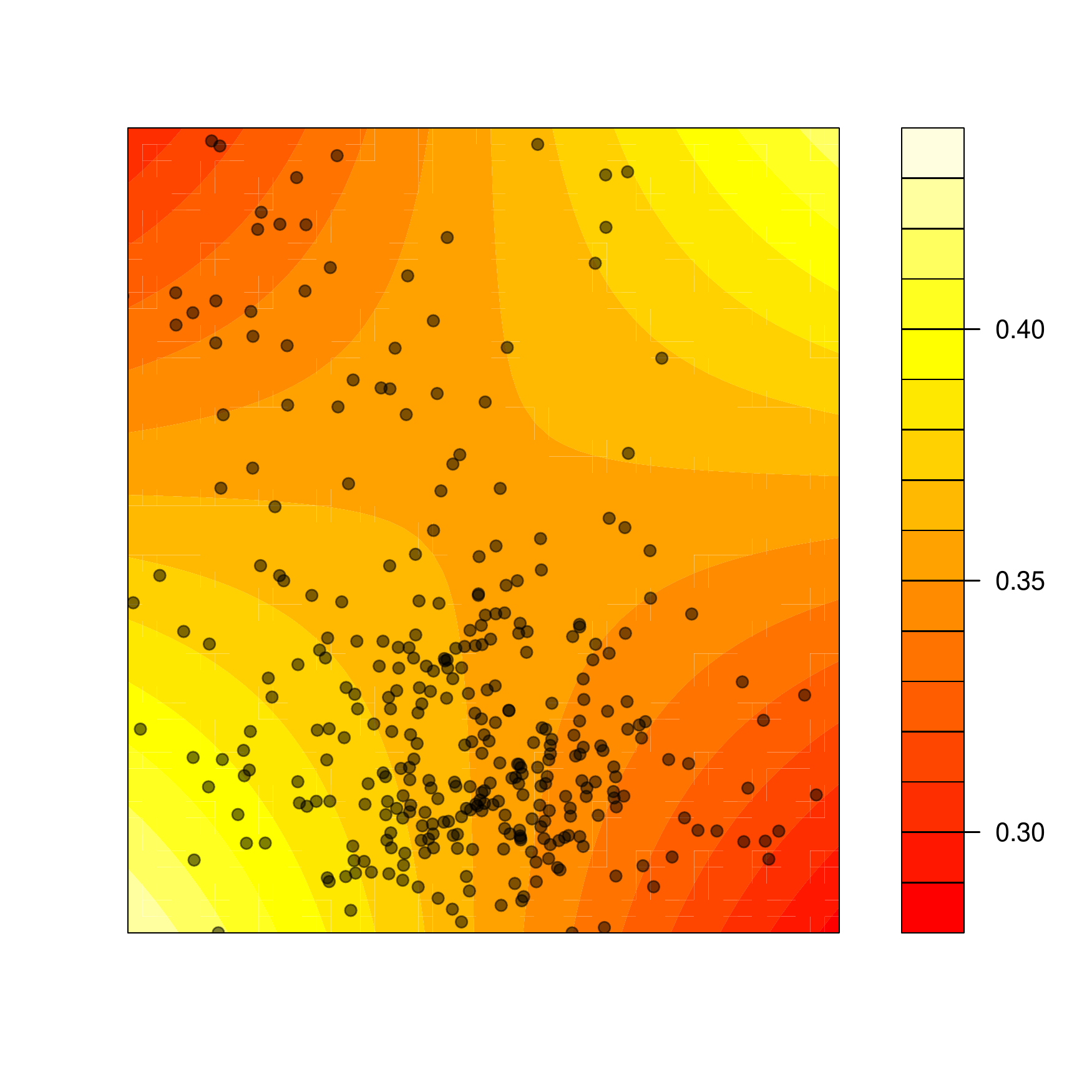}
  \caption{Mixture,  PC 1 vs {\tt mine} PC 1  }
\end{subfigure}
\begin{subfigure}{.33\textwidth}
  \centering
  \includegraphics[width=\linewidth]{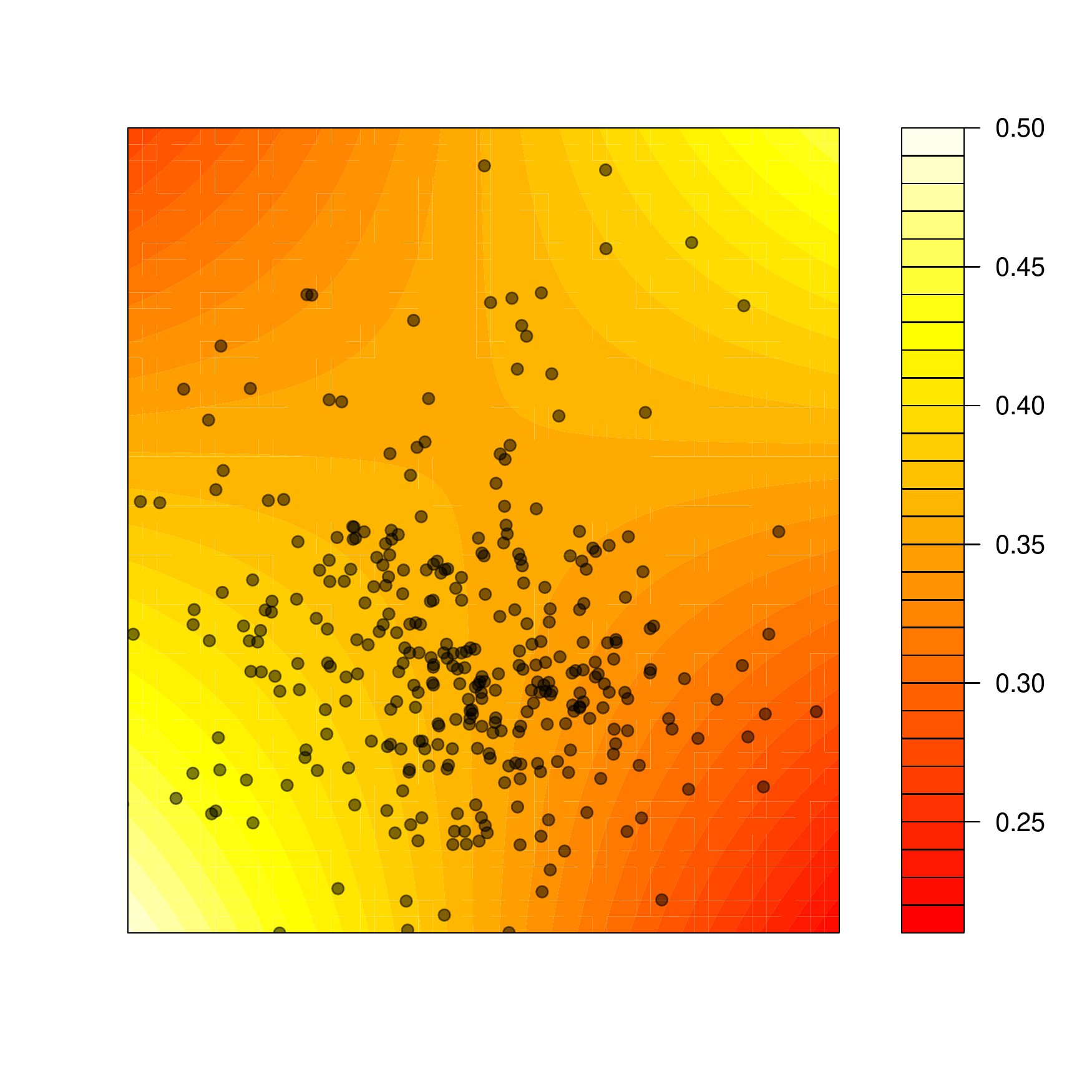}
  \caption{Mixture,  PC 1 vs {\tt mine} PC 3  }
\end{subfigure}
\begin{subfigure}{.33\textwidth}
  \centering
  \includegraphics[width=\linewidth]{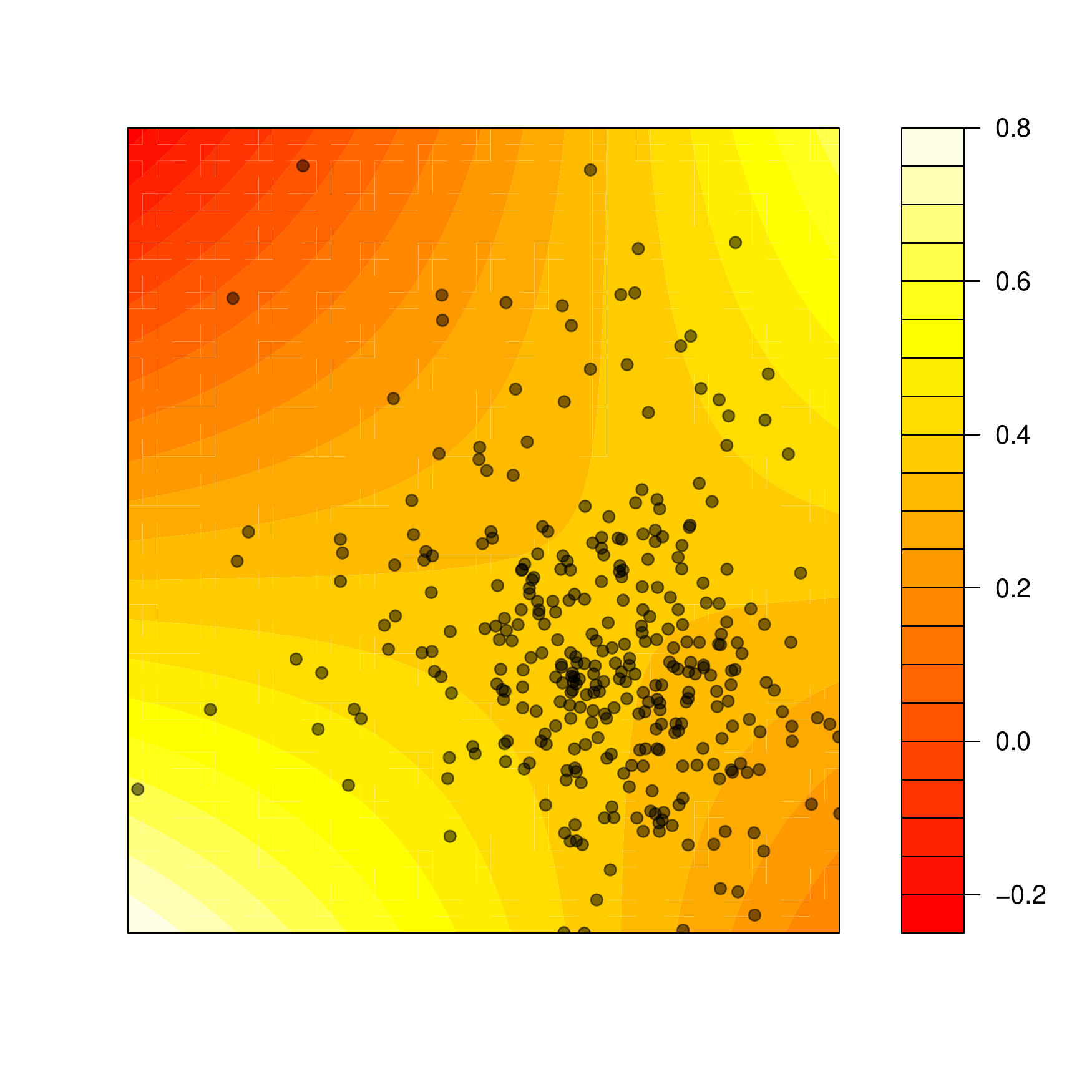}
  \caption{Mixture,  PC 3 vs {\tt mine} PC 3  }
\end{subfigure}

\begin{subfigure}{.33\textwidth}
  \centering
  \includegraphics[width=\linewidth]{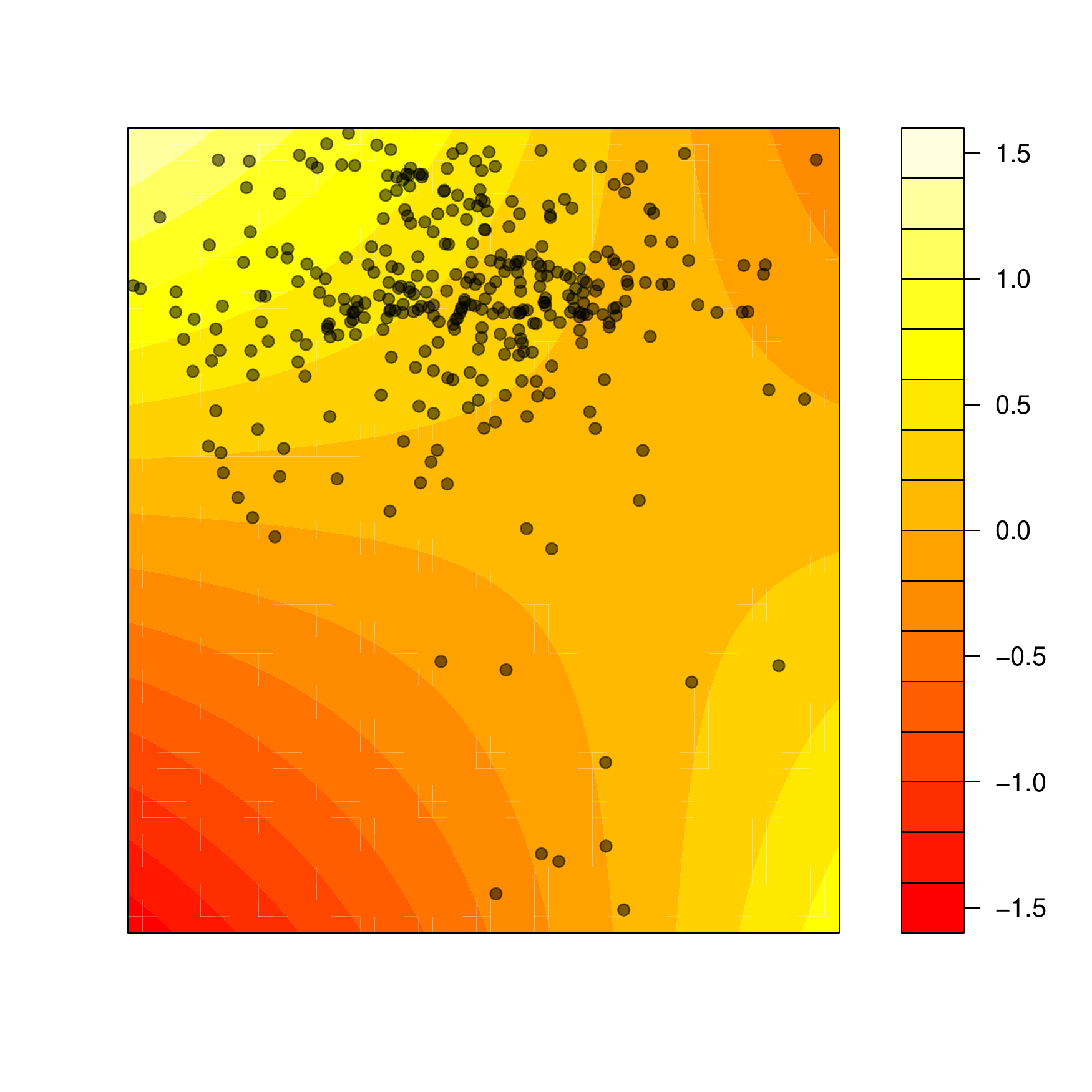}
  \caption{Mixture,  PC 1 vs {\tt vitA} PC 2  }
\end{subfigure}
\begin{subfigure}{.33\textwidth}
  \centering
  \includegraphics[width=\linewidth]{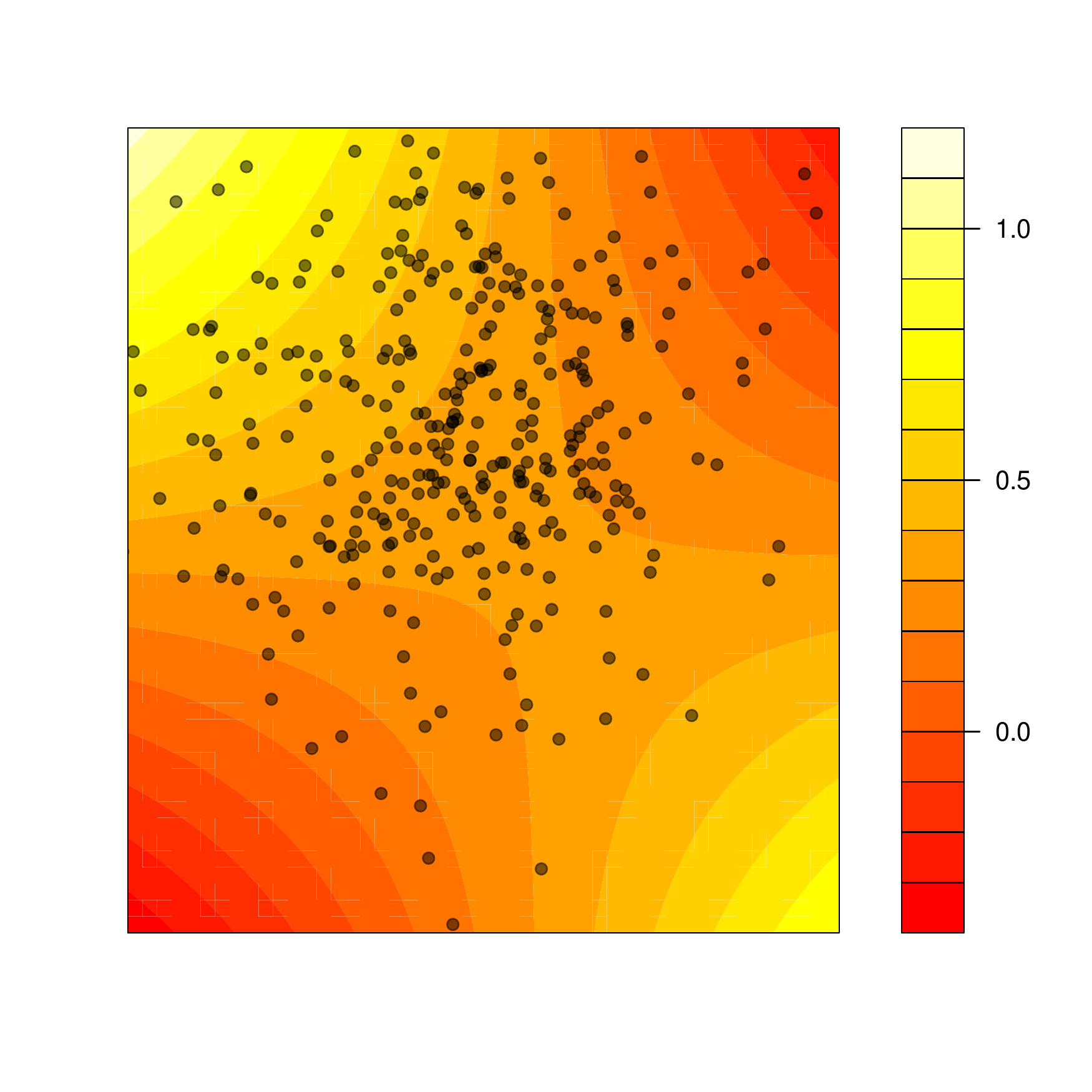}
  \caption{Mixture,  PC 1 vs {\tt vitB} PC 1  }
\end{subfigure}
\begin{subfigure}{.33\textwidth}
  \centering
  \includegraphics[width=\linewidth]{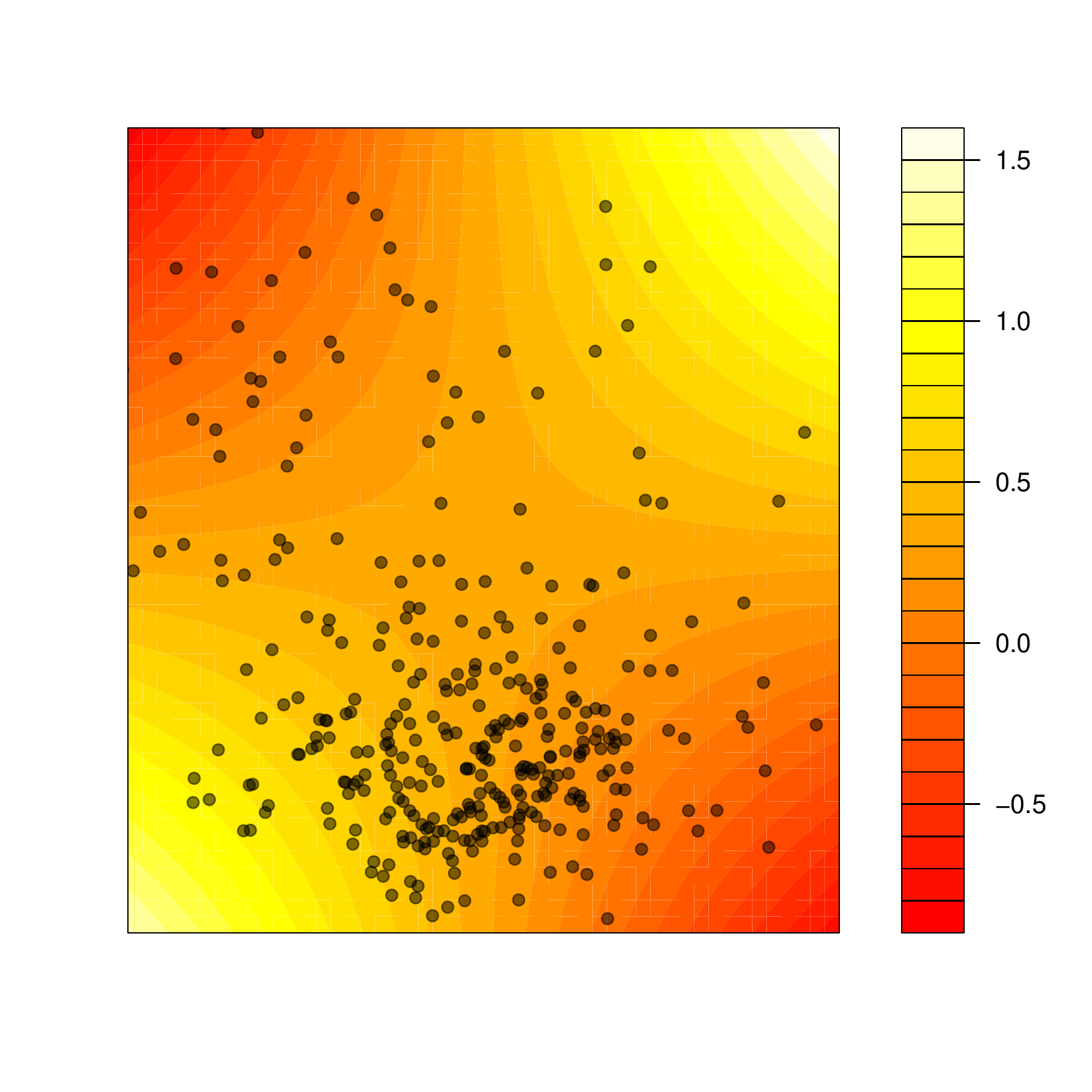}
  \caption{Mixture,  PC 1 vs {\tt vitB} PC 2  }
\end{subfigure}

\begin{subfigure}{.33\textwidth}
  \centering
  \includegraphics[width=\linewidth]{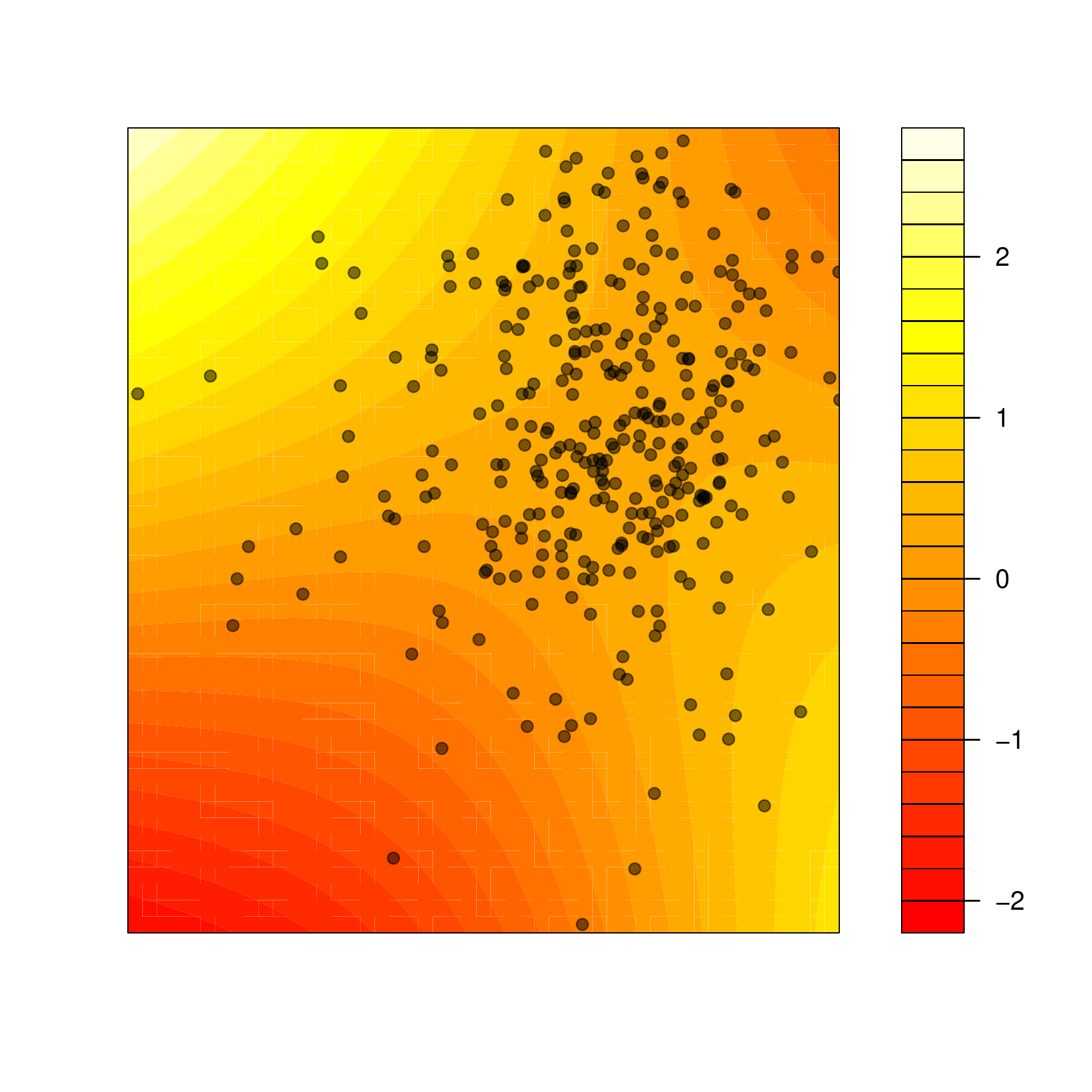}
  \caption{Mixture,  PC 3 vs {\tt vitB} PC 1  }
\end{subfigure}
\begin{subfigure}{.33\textwidth}
  \centering
  \includegraphics[width=\linewidth]{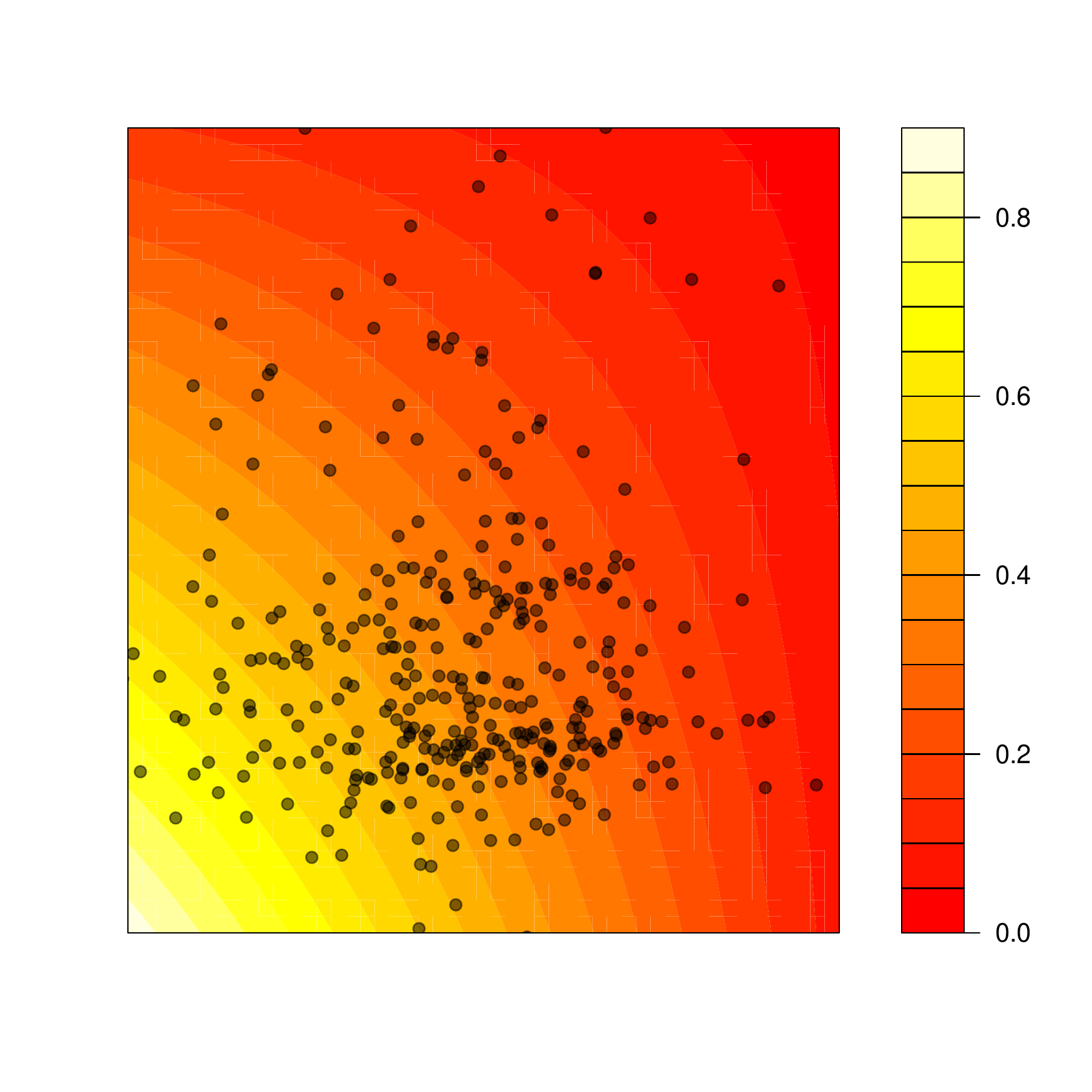}
  \caption{Mixture,  PC 1 vs {\tt vitO} PC 2  }
\end{subfigure}
\begin{subfigure}{.33\textwidth}
  \centering
  \includegraphics[width=\linewidth]{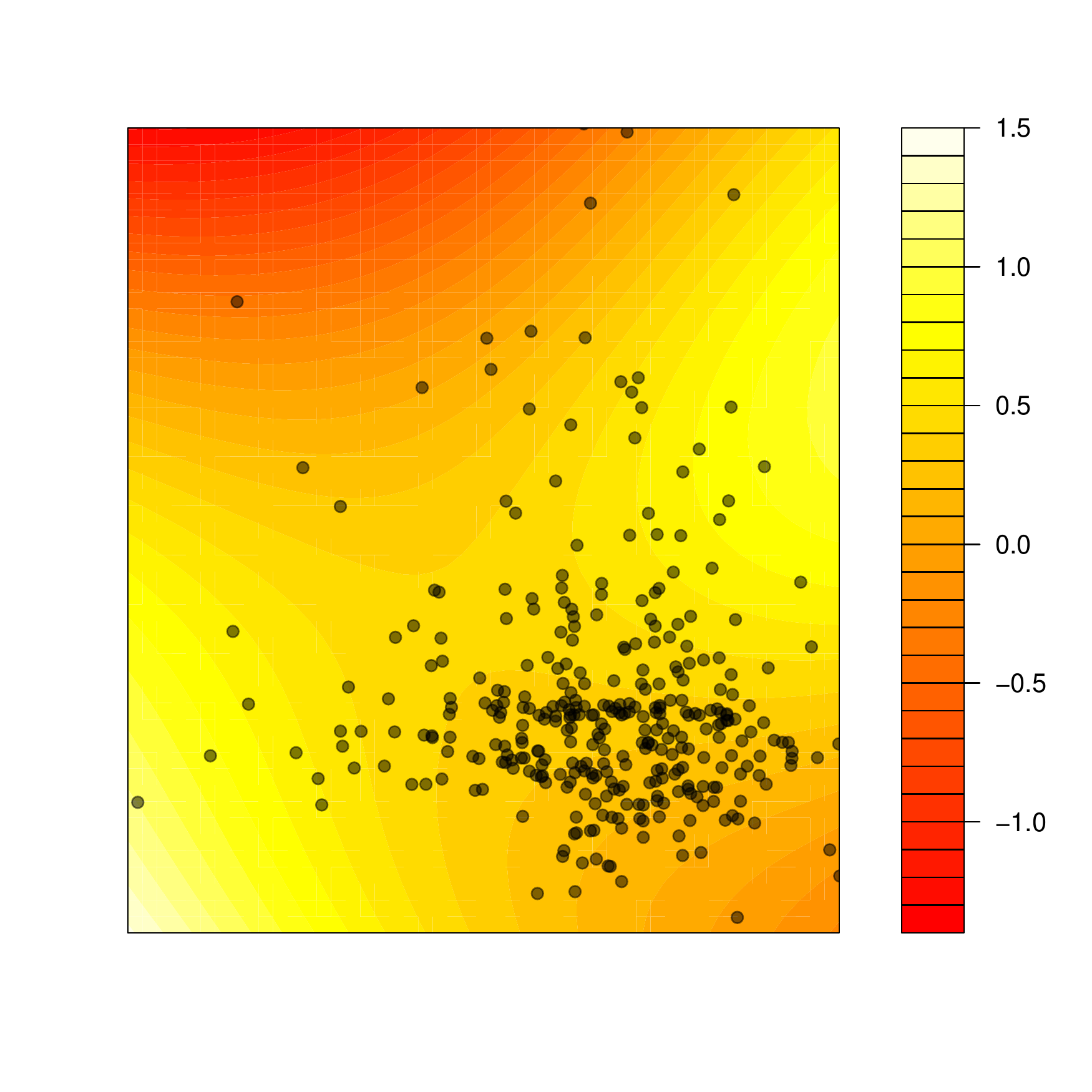}
  \caption{Mixture,  PC 3 vs {\tt vitO} PC 1  }
\end{subfigure}

\caption{Interactions between joint mixture and selected principal components in other four nutrition groups (i.e. Mineral, Vitamin A, Vitamin  B and Other Vitamins).}
\label{fig:int_other}
\end{figure}

To better understand the nature of the multivariate interactions between the environmental exposures and nutrition,  in Figure \ref{fig:int_macro} and \ref{fig:int_other}, we visualize the fitted exposure-response surface relating the mean normalized fine motor BSID-III score and the principal components (PCs) of the pollution mixture and of the nutrient groups.  Every panel in Figure \ref{fig:int_macro} and \ref{fig:int_other} depicts the joint effect of a pollutant PC and a nutrient PC for a selected nutrient group on the fine motor score, holding all the other PCs at their median. For each joint-effect term, the strength of evidence of an interaction between metal exposure and nutrition is driven by the "importance" of the corresponding PCs, i.e. the amount of variation the corresponding PCs explain in their respective pollutant/nutrient group. For example, in Figure \ref{fig:int_macro}, the pollutant PCs account for $42.6 \%$, $37.3\%$ and $20.1\%$ of the total variation in pollutant mixture, and the nutrient PCs account for $63.5 \%$, $28.5 \%$ and $7.4\%$ of the total variation in the macronutrient group. Consequently, the strength of the signal of the interaction between the $1^{st}$ PCs (e.g. Figure \ref{fig:int_macro} (a)) in the overall interaction is expectedly much stronger than that between the $3^{rd}$ PCs (e.g. Figure \ref{fig:int_macro} (i)). This explains the lack of significant evidence of overall interaction for the macronutrients in Table \ref{tb:res_joint}, since the joint effect between the PCs accounting for more variance (e.g. Figure \ref{fig:int_macro} (a),(b) and (d)) do not display strong evidence of  interaction. In comparison, for the other four nutrient groups, evidence of interaction can be observed between at least two nutrient-pollutant PC pairs among their leading PCs (Figure \ref{fig:int_other}), thereby suggesting evidence of overall interaction between nutrients and the pollutant mixture, and consequently providing additional evidence for the findings from the CVEK tests in Table \ref{tb:res_joint}. Finally, we observe that across all nutrient groups, the nutrient PCs interacts the most often with the $1^{st}$ pollutant PC, which is strongly associated with As, and also with the $3^{rd}$ pollutant PC, which is strongly associated with Mn, suggesting that As and Mn are the two main pollutants driving the overall interaction. Furthermore, the pattern of interaction between nutrient and pollutant PCs are observed to be similar across nutrient groups: at lower levels of nutrients (x-axis), higher levels of metal exposure (y-axis) is associated with lower neurodevelopment scores. At intermediate or high levels of the nutrient, however, this negative association either disappears (see, e.g. Figure \ref{fig:int_other} (c), (d), (h)) or even becomes positive  (see, e.g. Figure \ref{fig:int_other} (a), (b),  (f)).

\section{Discussion}

Under the framework of kernel machine regression, we have developed a hypothesis testing procedure for detecting nonlinear interactions between groups of continuous covariates. In this context,  we identified the unique challenge of possible kernel misspecification for the main-effect terms in the model, and illustrated the negative consequences of misspecified main effect kernels both in terms of Type I error and power.  Specifically, we showed that an overly smooth model, even when including all causal covariates, can still underfit the data under the null and thereby produce inflated Type I error rates. On the other hand, an overly flexible model tends to overfit the data under both the null and the alternative, resulting in deflated Type I error and weak power. While these observations motivate careful selection of the form of the main effect kernels, we also observe that choice of regularization parameters via a likelihood-based model selection strategy  (for example, estimating the bandwidth parameter in a Gaussian RBF kernel via REML \cite{maity_powerful_2011}) can also over-smooth the main-effect terms under the null. This situation appears to be especially severe in limited sample sizes and for misspecified kernel functions (Figure \ref{fig:res} (a)-(c)). Our work addresses this challenge by  estimating the main-effect model using a flexible ensemble of carefully selected base kernels, which we term Cross-validated Ensemble of Kernels (CVEK), 
coupled with a hyperparameter selection strategy based on cross validation. This approach avoids kernel misspecification under the null and mitigates overfitting under the alternative,  resulting in tests that are powerful yet maintain nominal Type I error rates.  We validated the approach through  extensive numerical studies. Under a wide variety of data-generation mechanisms, CVEK consistently produced correct Type I error and reasonable power.

We applied the proposed method to estimate nutrition-environment interactions between exposure to a metal mixture and multiple nutrient groups on neurodevelopment in Bangladeshi children. Challenges presented by the analysis included the presence of nonlinear within-group interactions within the effect of the metal mixture, the high-dimensionality for the between-group interaction terms ($d_{N \times E} \geq 9 $), and the limited sample size ($n = 351$). 
The proposed test identified evidence of interaction between the metal mixture and four nutrient groups, and we observed differences between the CVEK-based results and those from  existing approaches for the mineral group. 
Visualization of bivariate exposure-response surfaces based on nutrient and metal PCs allowed us to visualize the direction of these interactions.   
The application is important in that identification of nutritional factors that can effectively mitigate the impact of adverse effects of environmental exposures can inform  recommendations for pregnant women to improve the health of children across the lifespan. 

A natural extension of the proposed method would be to apply variable selection to identify the most important subsets of exposures driving the detected multivariate interaction. One possibility would be through use of an Automatic Relevance Determination (ARD)-type approach, putting sparse-inducing constraints for each variable in the interaction kernel to prune out the effect of irrelevant exposures \cite{duvenaud_structure_2013}. Furthermore, the ensemble weights $\{u_d\}_{d=1}^D$ (see (\ref{eq:cvek})) in CVEK were estimated to maximize the estimator's cross-validation stability. The optimality of such method in terms of the power of the hypothesis test has not been fully investigated. It is desirable to develop an optimal estimation procedure for the ensemble weights $\{u_d\}_{d=1}^D$ that maximizes the power of the hypothesis test, in a manner similar to \cite{gretton_optimal_2012}. Given such a procedure, it is also of theoretical interest to compare the difference between the ensemble weights generated by maximizing cross-validation stability to those generated by maximizing the power of the test in both finite samples and asymptotically. 
Finally, due to the abundance of non-Gaussian and non-continuous outcomes in the epidemiology and medical literature, extension of the kernel ensemble approach to discrete or censored continuous outcomes is also of interest.
\newpage

\section*{Tables and Figures}

\begin{algorithm}
\caption{Cross Validated Kernel Ensemble (CVKE)} 
\label{alg:cvke}
\begin{algorithmic}[1]
\Procedure{CVKE}{}\newline
\textbf{Input:} A library of kernels $\{k_d\}_{d=1}^D$, Data $(\by, \bX, \bx)$\newline
\textbf{Output:} Ensemble Kernel Matrix $\widehat{\bK}$\newline
$\#$ \texttt{Stage 1: Estimate $\lambda$ and CV error for each kernel}
\For{$d = 1$ to $D$}
\State{$\bK_d = \bK_d/tr(\bK_d)$}
\State{$\widehat{\lambda}_d = \mbox{\textit{argmin}} \; 
\texttt{CV}\Big(\lambda | \bK_d \Big)$} 
\State{$\widehat{\epsilon}_d = 
\texttt{CV}\Big(\widehat{\lambda}_d | \bK_d \Big)$}
\EndFor 
\newline
$\#$ \texttt{Stage 2: Estimate ensemble weights $\bu_{D \times 1} = \{u_1, \dots, u_D\}$}
\State{
$\widehat{\bu} = \underset{\bu \in \Delta}{argmin} \; ||\sum_{d=1}^D u_d \widehat{\epsilon}_d||^2 \qquad 
\mbox{where} \quad
\Delta = \{\bu| \bu \geq 0, ||\bu||_2^2 = 1 \}$}
\newline
$\#$ \texttt{Stage 3: Assemble the ensemble kernel matrix $\widehat{\bK}_{ens}$}
\State{$\widehat{\bA} = \sum_{d=1}^D 
\widehat{\mu}_d\bA_{\widehat{\lambda}_d, k_d}$}
\State{$\bU_A, \bdelta_{A} = 
\texttt{spectral\_decomp}(\widehat{\bA})$}
\State{$\lambda_{\bK} = min \Big(1,  
(\sum_{k=1}^n \frac{\delta_{A, k}}{1 - \delta_{A,k}})^{-1}, 
min \big(\{\widehat{\lambda}_d\}_{d=1}^D \big)
\Big)$}
\State{$\widehat{\bK} = 
\lambda_{\bK} * 
\widehat{\bU}_A \; 
diag\Big( \frac{\delta_{A,k}}{1 - \delta_{A,k}} \Big) \;
\widehat{\bU}_A^T$}
\EndProcedure
\end{algorithmic}
\end{algorithm}

\begin{table}[ht]
\centering
\resizebox{\columnwidth}{!}{%
\begin{tabular}{|ccll|}
\hline\hline
Kernel Family & Kernel Function & Model Name & Parameter Value \\
\hline\hline
\multirow{2}{*}{Polynomial} & \multirow{2}{*}{$(1 + \bx^T\bx')^d$} & 
Linear & $d=1$\\
& & Quadratic & $d=2$ 
\\ \hline
\multirow{2}{*}{Gaussian RBF} & \multirow{2}{*}{$exp(-||\bx - \bx'||^2/\sigma^2)$} & 
RBF-MLE & $\sigma={\tt argmax}(\mbox{REML}(\sigma))$\\
& & RBF-Median & $\sigma={\tt median}(\{||\bx_i - \bx_j||\}_{i \neq j})$ 
\\ \hline
\multirow{3}{*}{Mat\'{e}rn} & 
\multirow{3}{*}{$\frac{2^{1 - \nu}}{\Gamma(\nu)}
\Big( \sqrt{2 \nu} \sigma ||\br||\Big)^\nu
K_{\nu} \Big( \sqrt{2 \nu} \sigma ||\br|| \Big)$} & 
Mat\'{e}rn 1/2 & $\nu=1/2$\\
& & Mat\'{e}rn 3/2 & $\nu=3/2$ \\
& & Mat\'{e}rn 5/2 & $\nu=5/2$ 
\\ \hline
\multirow{3}{*}{Neural Network} & 
\multirow{3}{*}{$\frac{2}{\pi} * sin^{-1} \Big(
\frac{2  \sigma\tilde{\bx}^T\tilde{\bx}'}
{\sqrt{(1 + 2  \sigma\tilde{\bx}^T \tilde{\bx})(1 + 2 \sigma \tilde{\bx}'^T\tilde{\bx}')}}\Big)$} & 
NN 0.1 & $\sigma=0.1$\\
& & NN 1 & $\sigma=1$ \\
& & NN 10 & $\sigma=10$ 
\\ \hline
\multirow{2}{*}{CVEK} & 
\multirow{2}{*}{
$\widehat{\bK} = \lambda_\bK * \Big[
\bU_A diag \Big( \frac{\delta_{A, k}}{1 - \delta_{A,k}} \Big) \bU_A^T \Big]$
} & 
CVEK-RBF & $log(\sigma) \in \{-2, -1, 0, 1, 2\}$\\
& & CVEK-NN & $\sigma \in \{0.1, 1, 10, 50\}$ \\
\hline\hline
\end{tabular}
}
\caption{List of $k_{model}$'s considered in the numeric study}
\label{tb:k_model}
\end{table}

\begin{table}[!htbp] 
\centering 
\begin{tabular}{|c|ccccc|} 
\hline \hline 
\multirow{2}{*}{Model} & \multicolumn{5}{c|}{Nutrient Group}  \\ 
\cline{2-6}
 & macro & mineral & vitamin A & vitamin B & vitamin, other  \\ 
\hline \hline
 CVEK-NN & 0.1591 & 0.0586 & 0.0406 & 0.0445 & 0.0476 \\ 
 CVEK-RBF & 0.1908 & 0.0595 & 0.0396 & 0.0399 & 0.0452 \\ 
\hline \hline 
 iSKAT & 0.2485 & 0.1074 & 0.0476 & 0.0675 & 0.0451 \\ 
 GKM & 0.1654 & 0.1075 & 0.0471 & 0.0680 & 0.0357 \\
 GE-spline & 0.3743 & 0.1368 & 0.0084 & 0.2134 & 0.4338 \\
\hline\hline
\end{tabular} 
\caption{$p-value$ for Nutrient - Environment interaction test with joint As, Pb, Mn mixture } 
\label{tb:res_joint}
\end{table}

\clearpage
\begin{figure}[ht]
\centering
\resizebox{0.8\linewidth}{!}{
\begin{subfigure}{.33\textwidth}
  \centering
  \includegraphics[width=\linewidth]{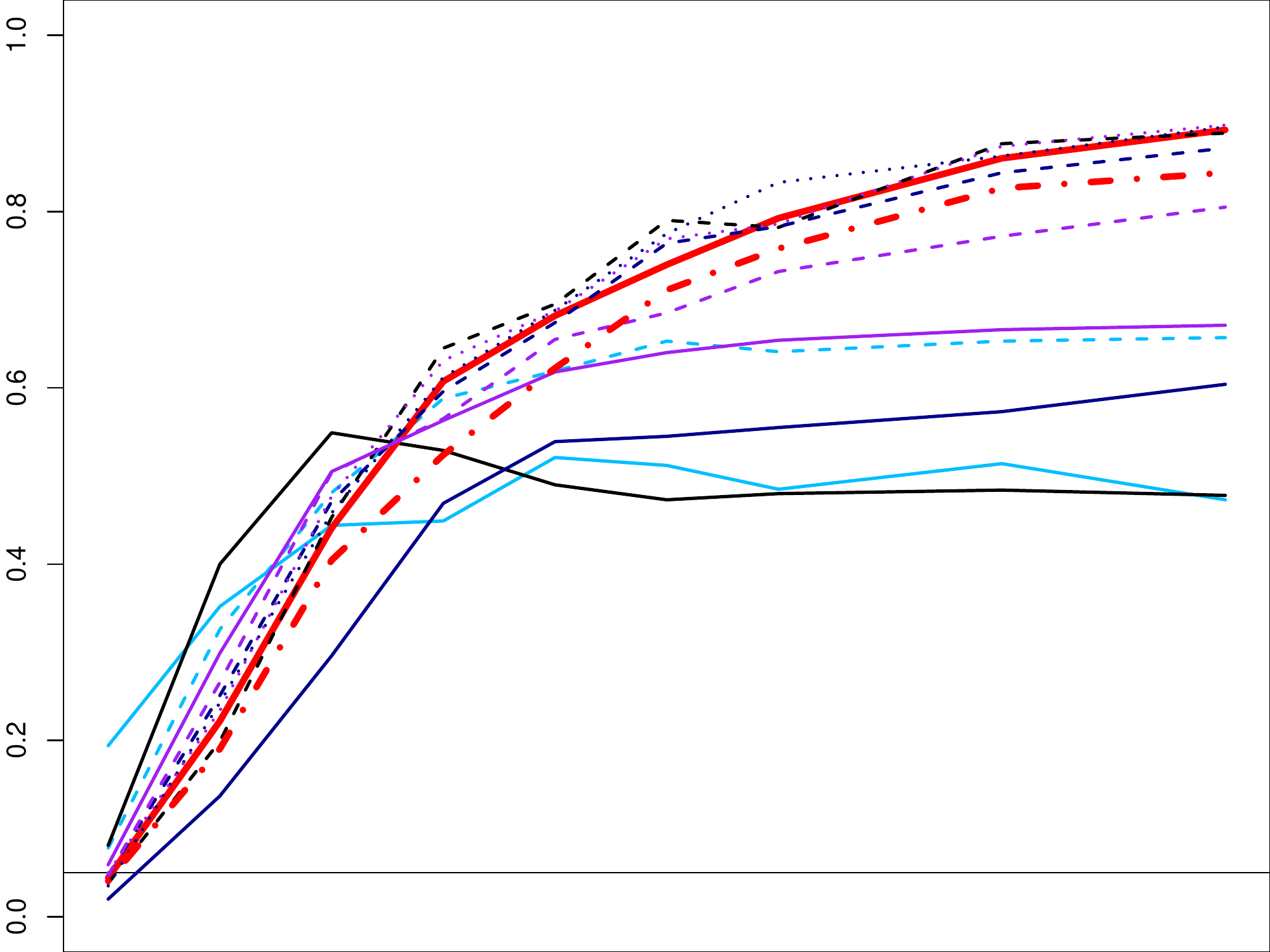}
  \caption{$k_{\texttt{true}}=$ Mat\'{e}rn 3/2, $\sigma = 0.5$}
  \label{fig:sfig11}
\end{subfigure}\hspace*{-0.1em}
\begin{subfigure}{.33\textwidth}
  \centering
  \includegraphics[width=\linewidth]{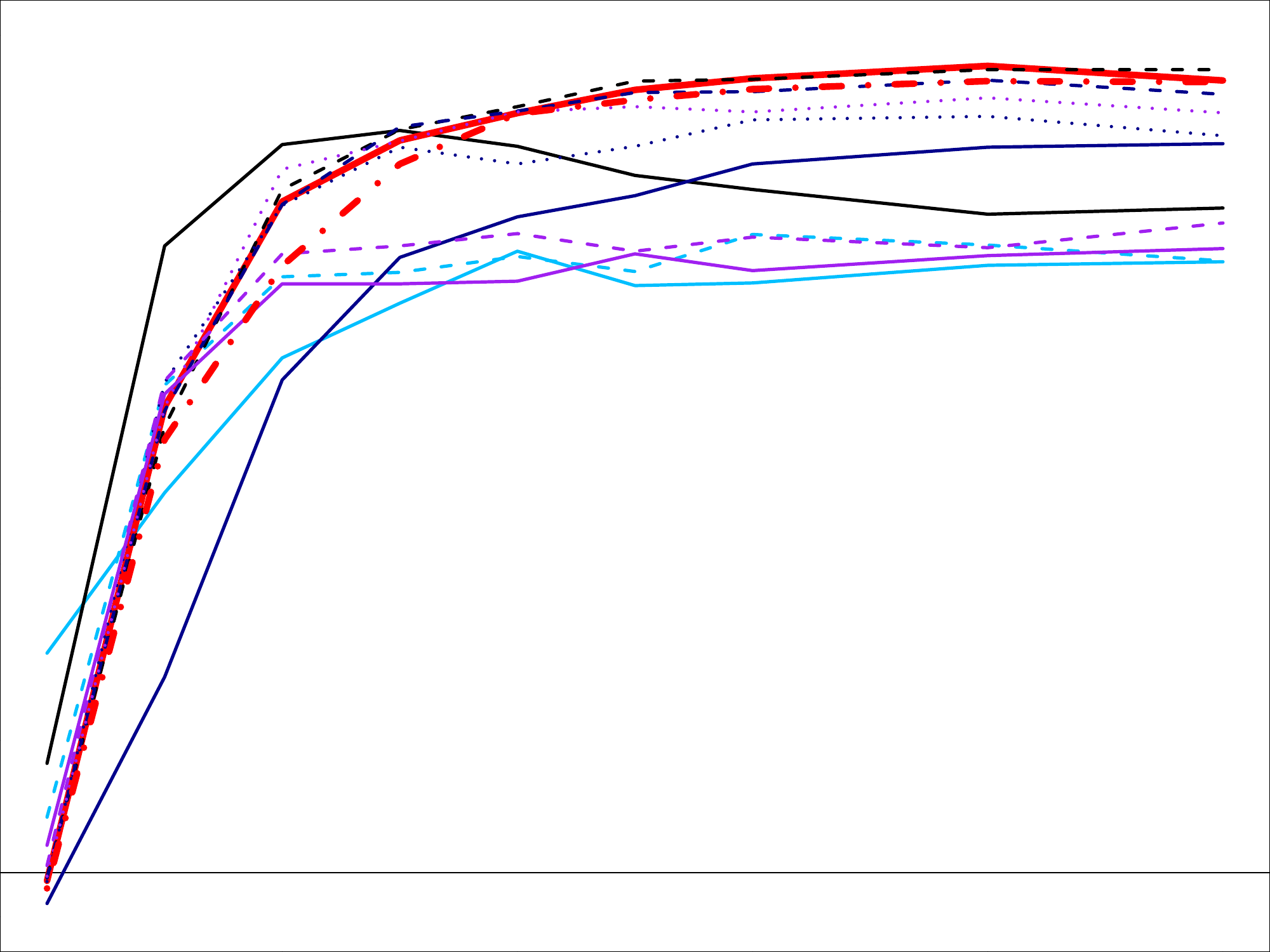}
  \caption{$k_{\texttt{true}}=$ Mat\'{e}rn 3/2, $\sigma = 1$}
  \label{fig:sfig12}
\end{subfigure}\hspace*{-0.1em}
\begin{subfigure}{.33\textwidth}
  \centering
  \includegraphics[width=\linewidth]{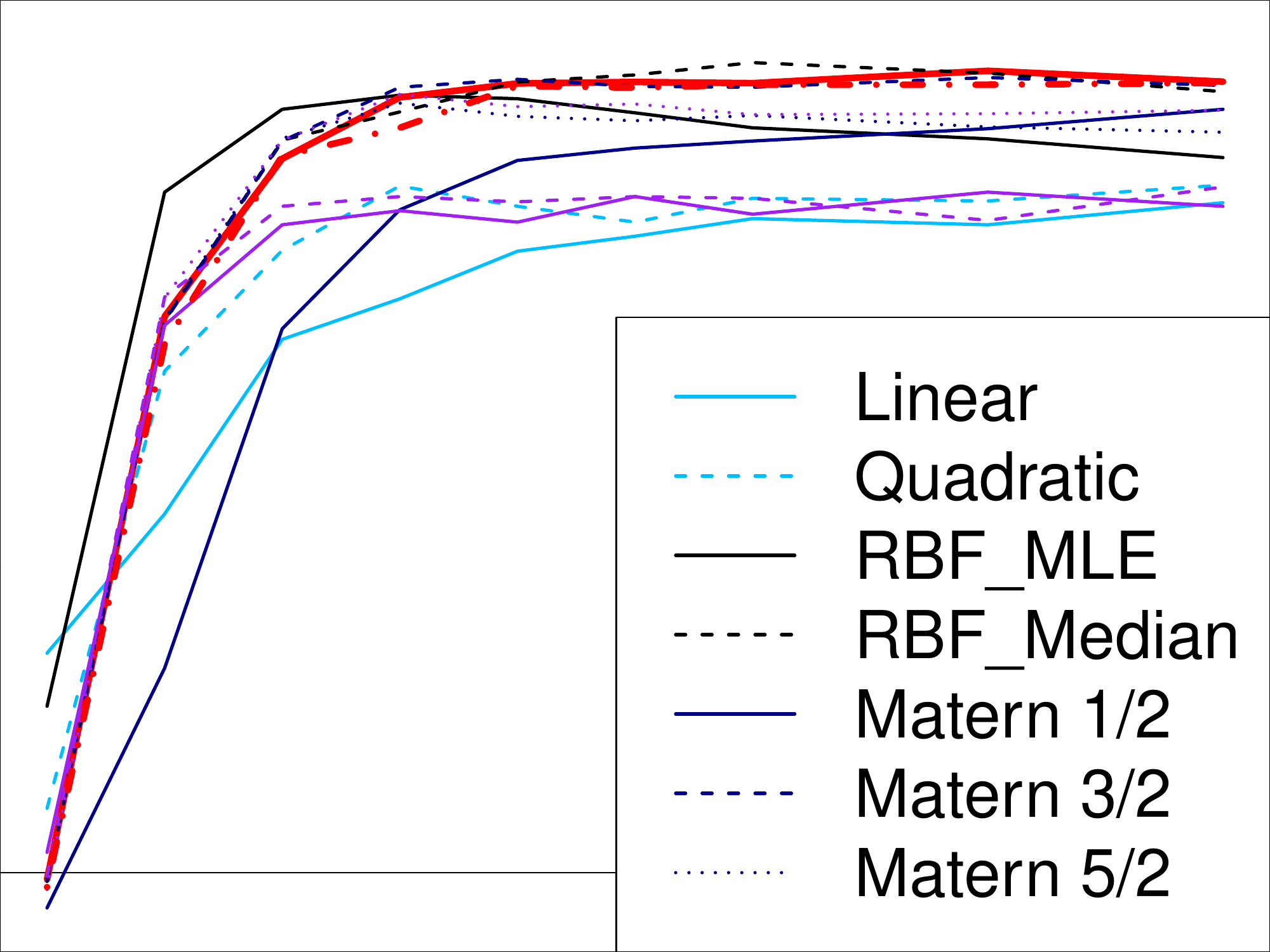}
  \caption{$k_{\texttt{true}}=$ Mat\'{e}rn 3/2, $\sigma = 1.5$}
  \label{fig:sfig13}
\end{subfigure}
}
\vspace*{-0.1em}
\resizebox{0.8\linewidth}{!}{
\begin{subfigure}{.33\textwidth}
  \centering
  \includegraphics[width=\linewidth]{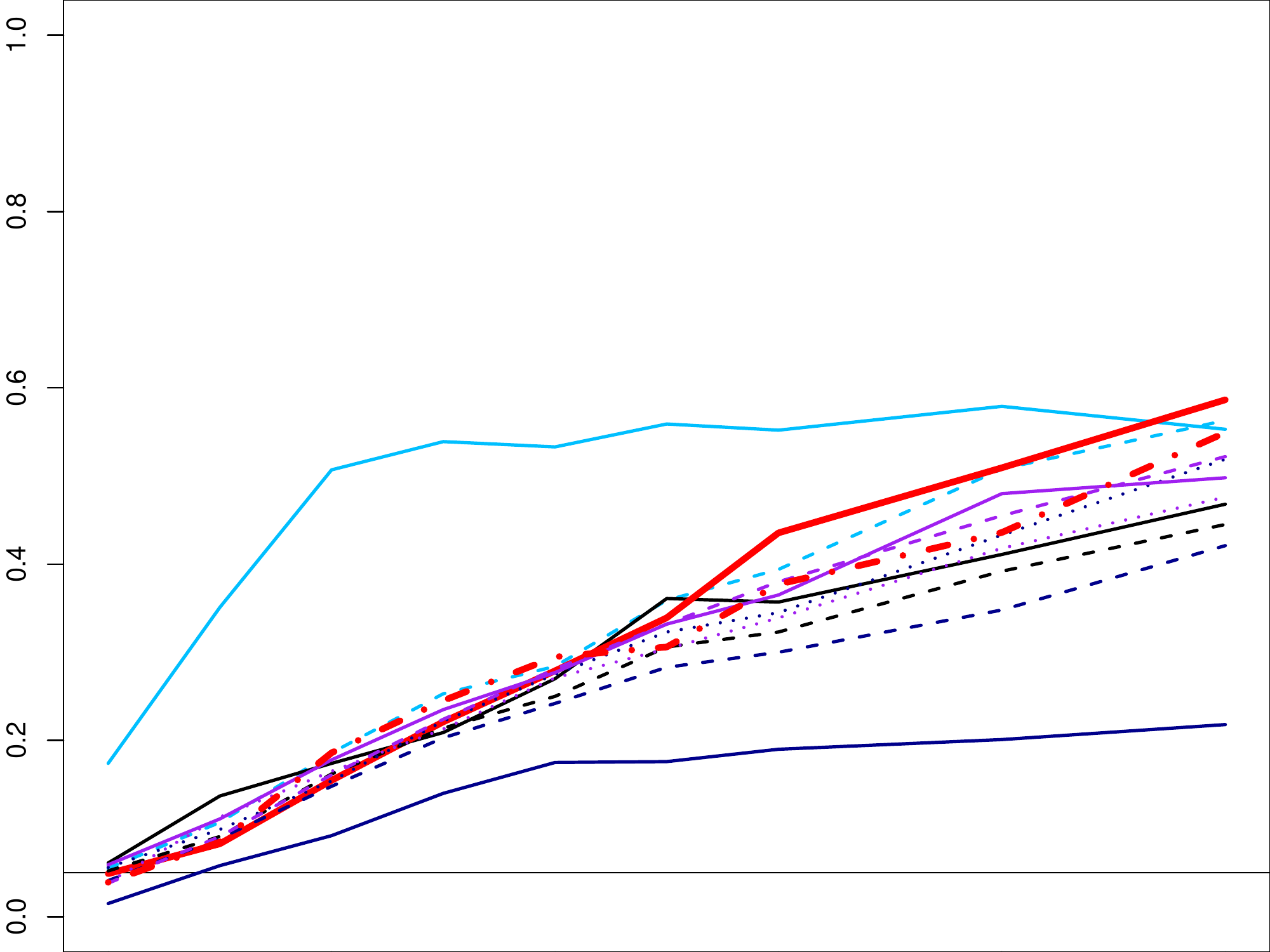}
  \caption{$k_{\texttt{true}}= $ Mat\'{e}rn 5/2, $\sigma = 0.5$}
  \label{fig:sfig21}
\end{subfigure}\hspace*{-0.1em}
\begin{subfigure}{.33\textwidth}
  \centering
  \includegraphics[width=\linewidth]{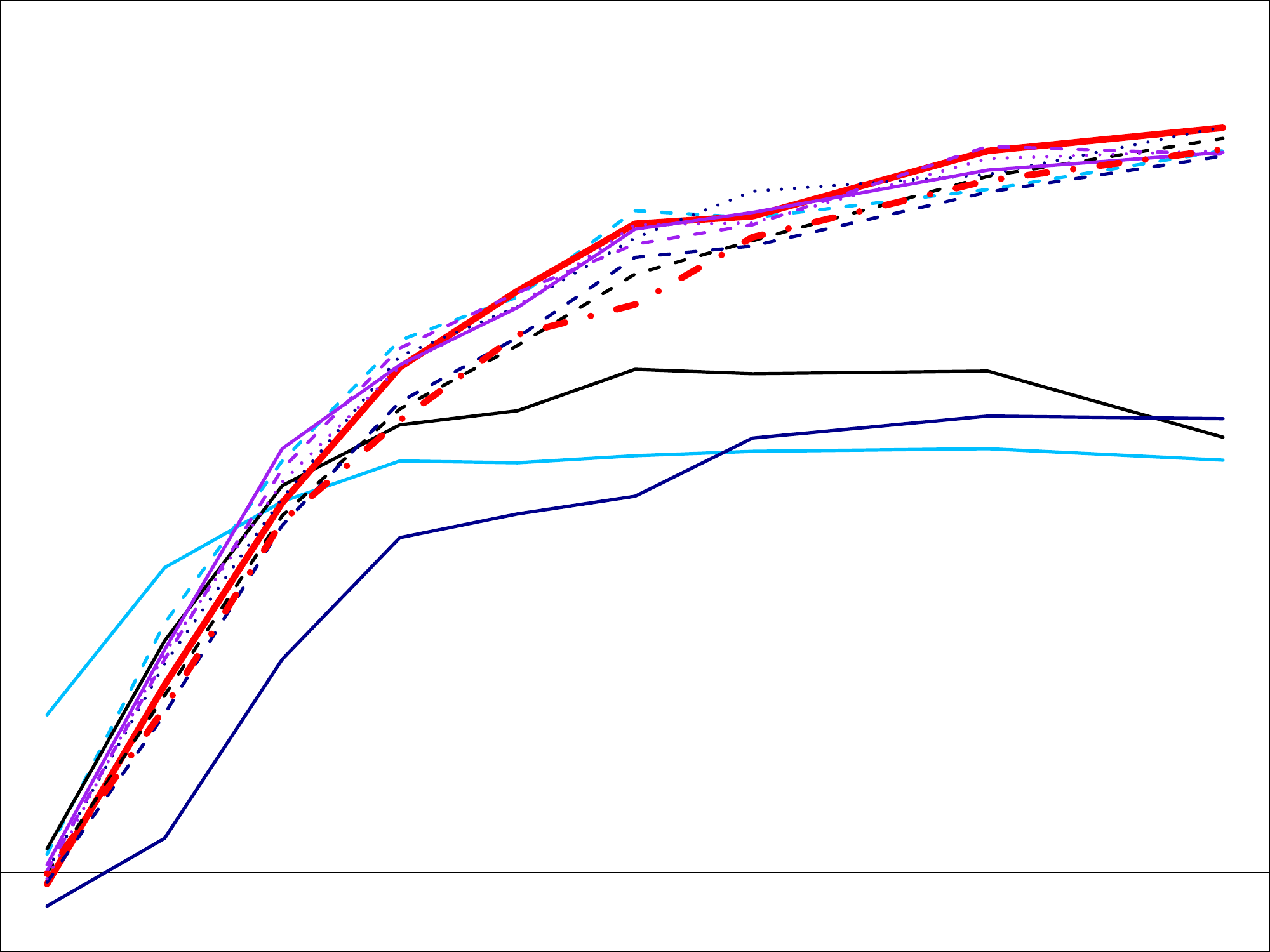}
  \caption{$k_{\texttt{true}}= $ Mat\'{e}rn 5/2, $\sigma = 1$}
  \label{fig:sfig22}
\end{subfigure}\hspace*{-0.1em}
\begin{subfigure}{.33\textwidth}
  \centering
  \includegraphics[width=\linewidth]{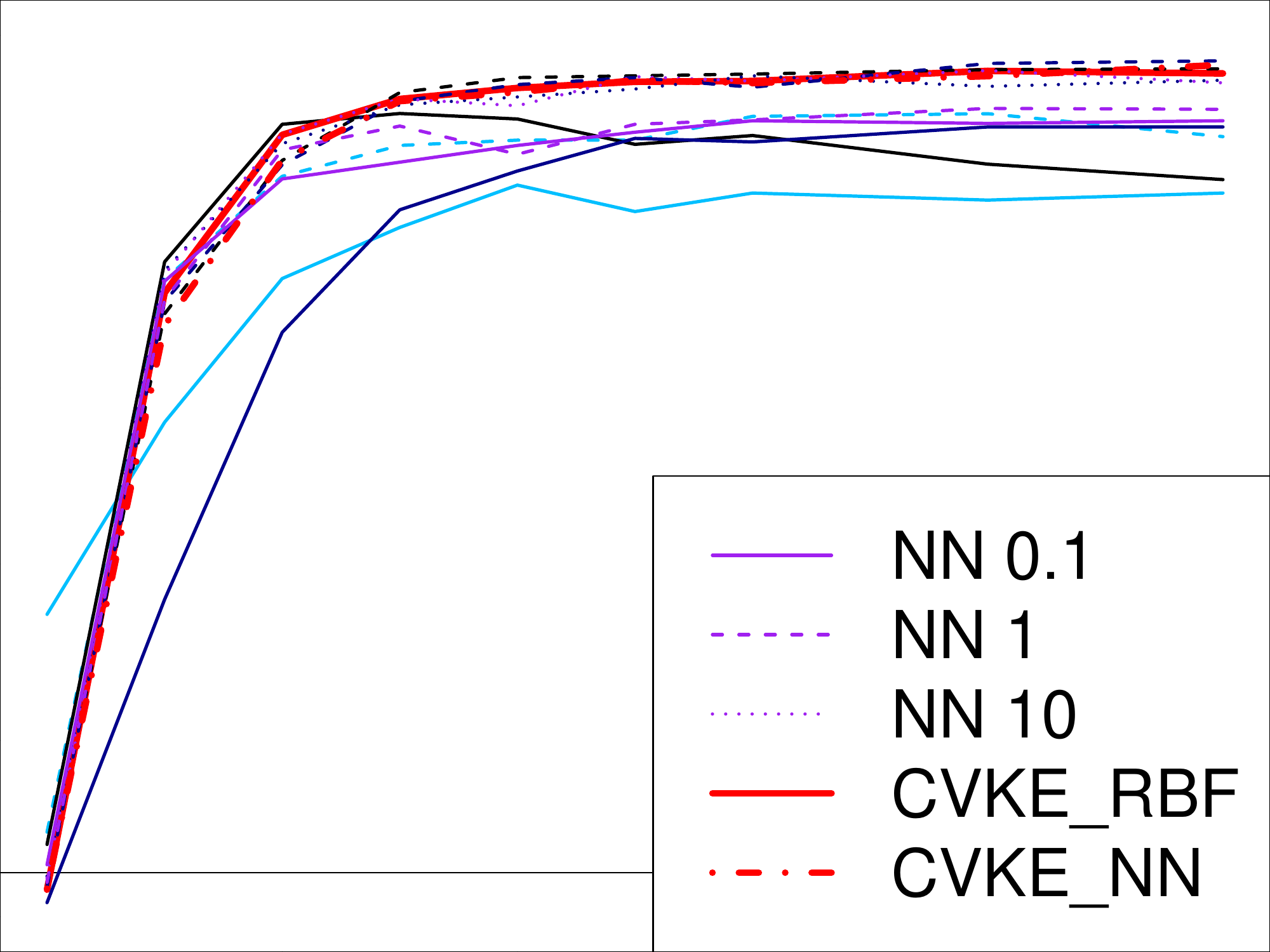}
  \caption{$k_{\texttt{true}}= $ Mat\'{e}rn 5/2, $\sigma = 1.5$}
  \label{fig:sfig23}
\end{subfigure}
}
\vspace*{-0.1em}

\resizebox{0.8\linewidth}{!}{
\begin{subfigure}{.33\textwidth}
  \centering
  \includegraphics[width=\linewidth]{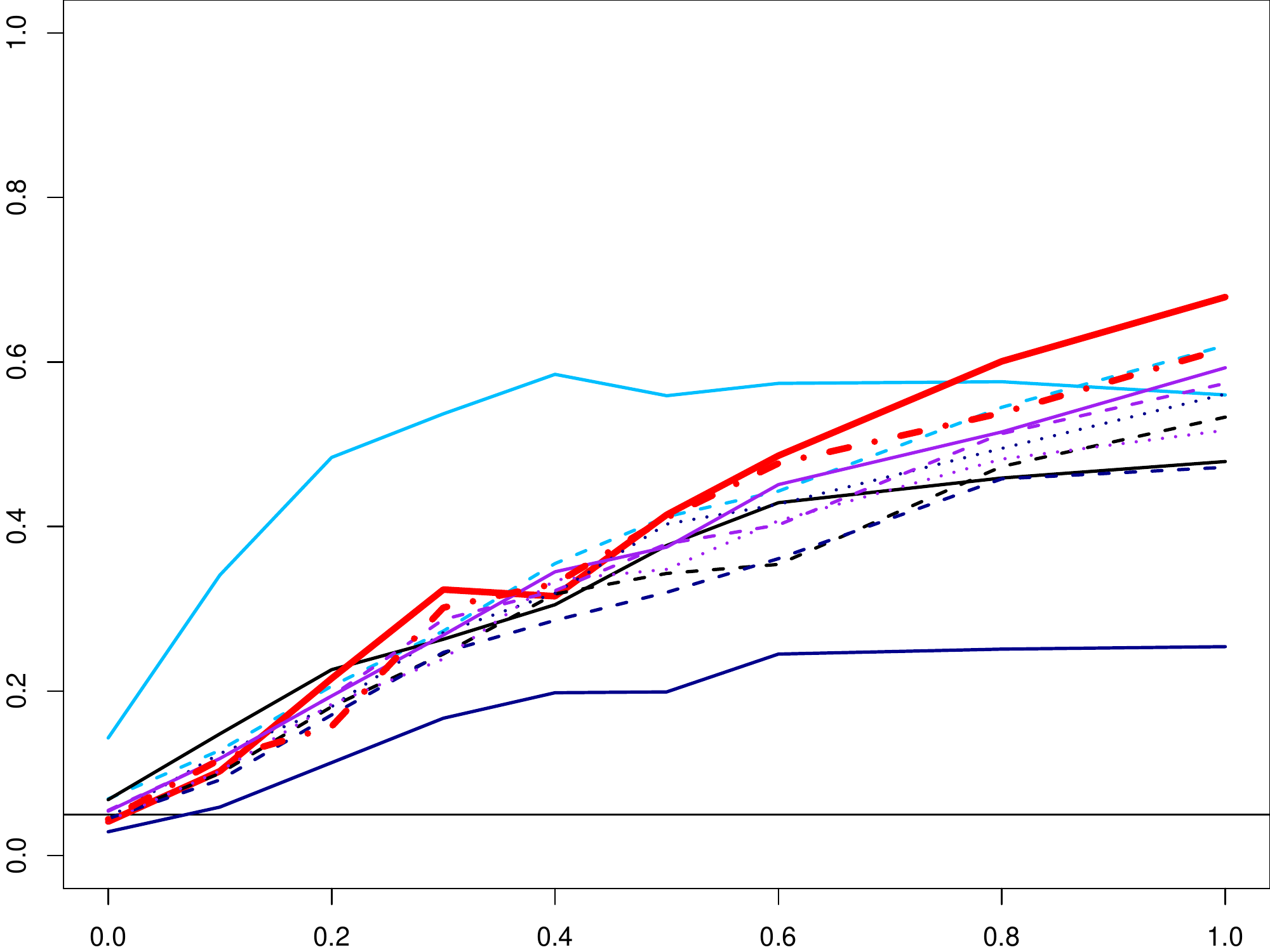}
  \caption{$k_{\texttt{true}}= $ Gaussian RBF, $\sigma = 0.5$}
  \label{fig:sfig31}
\end{subfigure}\hspace*{-0.1em}
\begin{subfigure}{.33\textwidth}
  \centering
  \includegraphics[width=\linewidth]{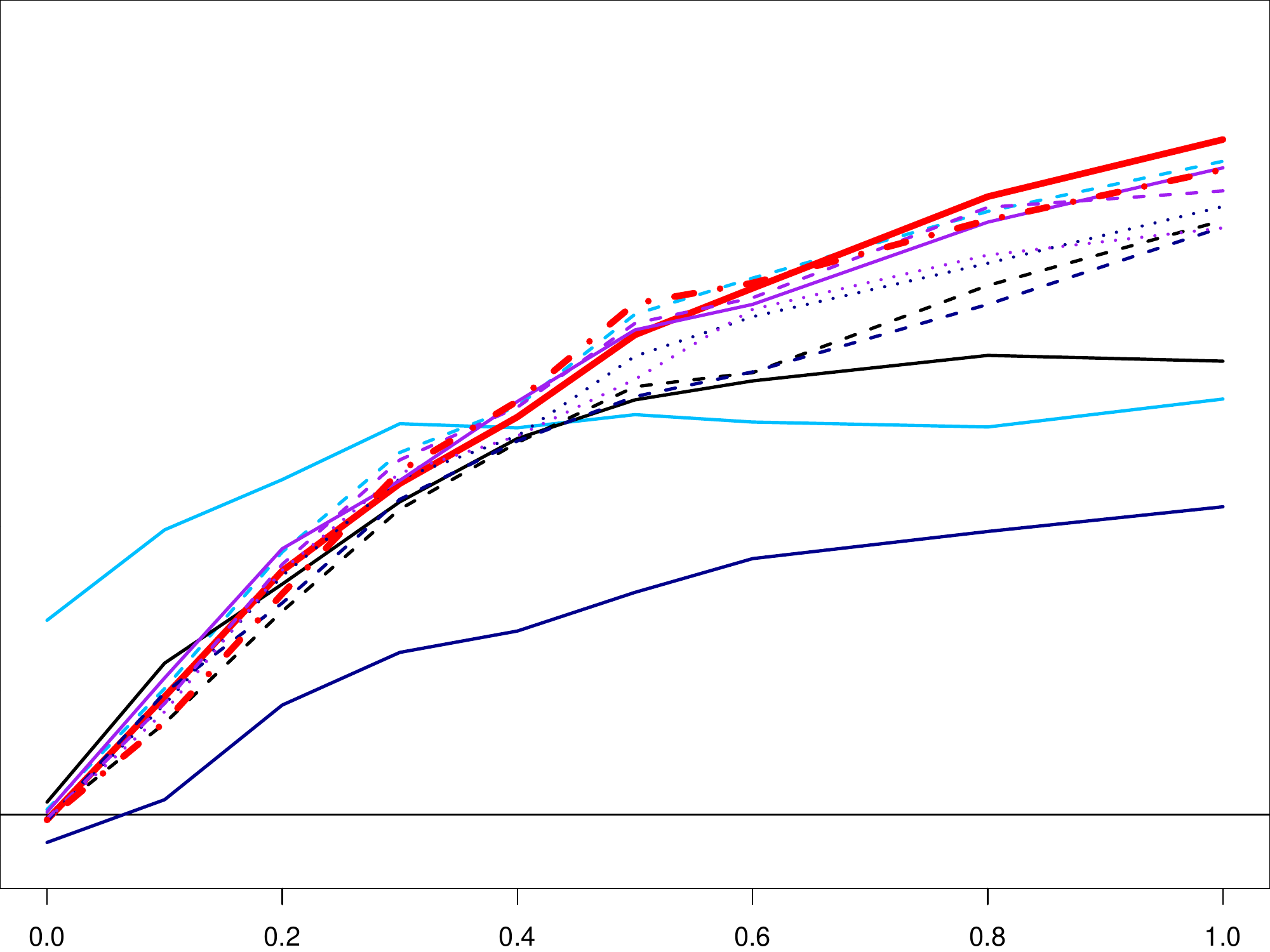}
  \caption{$k_{\texttt{true}}= $ Gaussian RBF, $\sigma = 1$}
  \label{fig:sfig32}
\end{subfigure}\hspace*{-0.1em}
\begin{subfigure}{.33\textwidth}
  \centering
  \includegraphics[width=\linewidth]{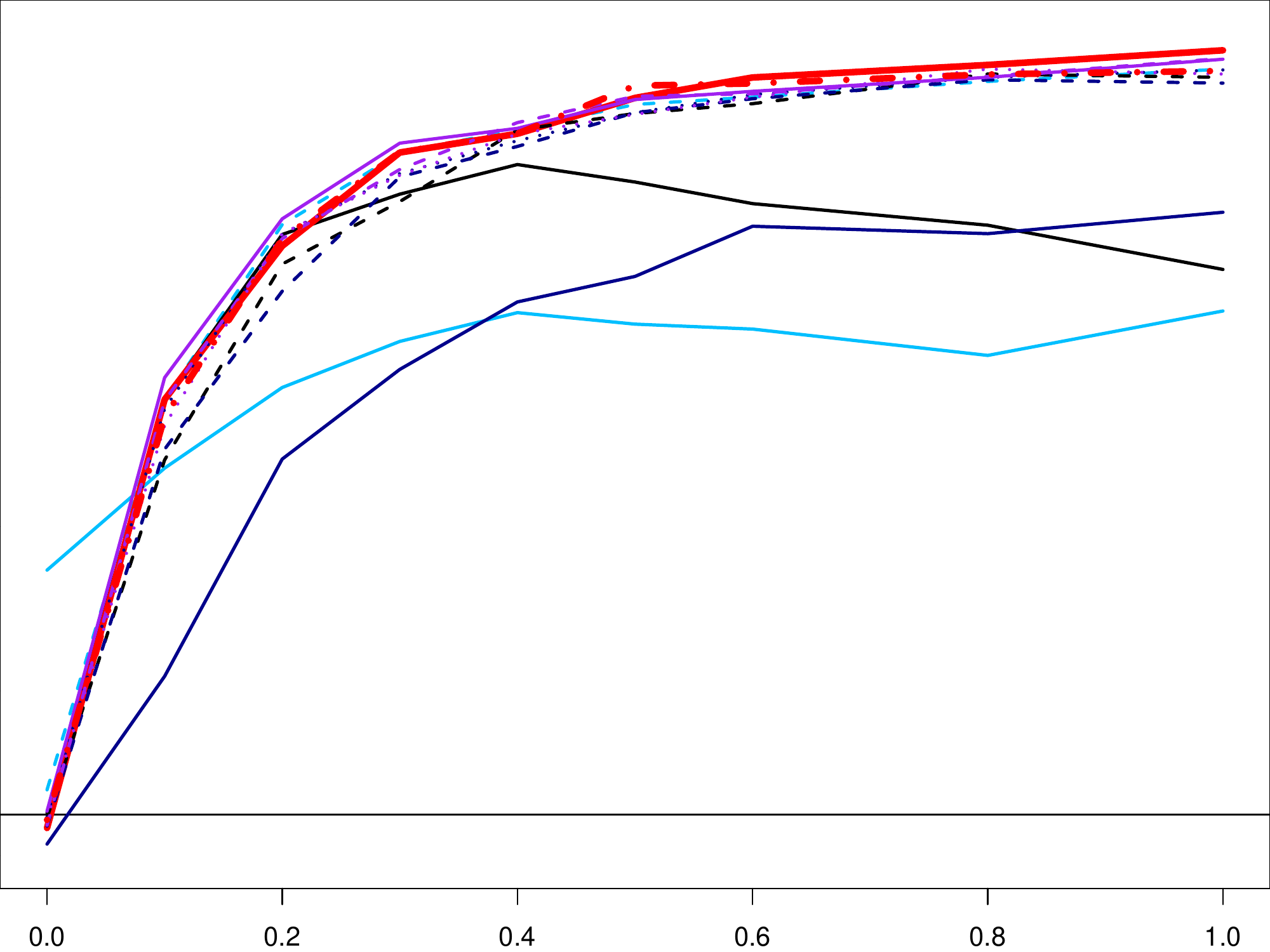}
  \caption{$k_{\texttt{true}}= $ Gaussian RBF, $\sigma = 1.5$}
  \label{fig:sfig33}
\end{subfigure}
}
\caption{Estimated $\widehat{P}(p < 0.05)$ (y-axis) as a function of Interaction Strength $\delta \in [0,1]$ (x-axis). \\
\textbf{Skype Blue}: Linear (Solid) and Quadratic (Dashed) Kernels, \textbf{Black}: RBF-Median (Solid) and RBF-MLE (Dashed), \textbf{Dark Blue}: Mat\'{e}rn Kernels with $\nu = \frac{1}{2}, \frac{3}{2}, \frac{5}{2}$, \textbf{Purple}: Neural Network Kernels with $\sigma = 0.1, 1, 10$, \textbf{Red}: CVEK based on RBF (Solid) and Neural Networks (Dashed). \\
Horizontal line marks the test's significance level (0.05). When $\delta = 0$, $\widehat{P}$ should be below this line.
}
\label{fig:res}
\end{figure}

\newpage

\clearpage
\bibliographystyle{abbrv}

\newpage

\appendix

\section{Appendix}
\renewcommand{\thefigure}{\thesection.\arabic{figure}}    
\renewcommand{\thetable}{\thesection.\arabic{table}} 
\setcounter{figure}{0}
\setcounter{table}{0}

\subsection{Derivation for Ensemble Kernel Matrix}
\label{sec:ekm_deriv}

Given the ensemble hat matrix $\widehat{\bA}$ in Section 4, we consider how to identify the ensemble kernel matrix $\widehat{\bK}$ by solving:
\begin{align*}
\widehat{\bK} (\widehat{\bK} + \lambda_\bK \bI)^{-1} = \widehat{\bA}.
\end{align*}

Specifically, if denote $(\bU_A, \bU_K)$ and $(\{\delta_{A,k}\}_{k=1}^n, \{\delta_{K,k}\}_{k=1}^n)$ the eigenvector and eigenvalues of $\widehat{\bA}$ and $\widehat{\bK}$, respectively, then the above system reduces to:
\begin{align*}
\bU_A diag \Big( \delta_{A,k} \Big) \bU_A^T = 
\bU_K diag \Big( \frac{\delta_{K,k}}{\delta_{K,k} + \lambda_\bK} \Big) \bU_K^T
\end{align*}
and adopts closed form solution $\bU_K = \bU_A$ and $\delta_{K,k} = \lambda_\bK \frac{\delta_{A, k}}{1 - \delta_{A,k}}$. Therefore the ensemble kernel matrix $\widehat{\bK}$ is estimated as:
\begin{align*}
\widehat{\bK} = \lambda_\bK * \bU_A diag \Big( \frac{\delta_{A, k}}{1 - \delta_{A,k}} \Big) \bU_A^T.
\end{align*}

\textbf{Choice of ensemble tunning parameter $\lambda_{\bK}$}

Notice that we have left the "ensemble tunning parameter" $\lambda_\bK$ unspecified. In practice, $\lambda_\bK$ serves only as a constant scaling factor for the kernel matrix $\bK$, whose exact value does not impact either the prediction or the p-value calculation, since both procedures are scale invariant with respect to the kernel matrix. Therefore it can be set to a value of our choice. One common choice for $\lambda_\bK$ is to set $\lambda_\bK = min\Big(1,  (\sum_{k=1}^n \frac{\delta_{A, k}}{1 - \delta_{A,k}})^{-1} \Big)$ such that $tr(\widehat{\bK}) \leq 1$, this is because the Rademacher complexity of the overall ensemble can be upper-bounded as a function of $tr(\widehat{\bK})$ \cite{lanckriet_learning_2004}. Another interesting choice is $\lambda_{\bK} = O\Big( min(\{\hat{\lambda}_d\}_{d=1}^D) \Big)$. Intuitively, this means the tunning parameter for ensemble kernel matrix $\widehat{\bK}$ should grow in the same rate as the tunning parameter for the best-peforming base kernel, as the ensemble kernel matrix is expected to perform as well or better than the best-performing base kernel. Further, 
such choice provides guarantee on the generalization performance of the ensemble by bounding the ensemble kernel matrix's decay rate of the tail sum of the eigenvalues \cite{mendelson_performance_2003}.

\subsection{Discussion on theoretical aspects of CVEK}
\label{sec:theory}

\subsubsection{Oracle Selection}

We remind readers that the CVEK's ensemble strategy (Step 2) is similar to that of Jackknife Model Averaging (JMA) \cite{hansen_jackknife_2012} in the case of leave-one-out cross validation and the Super Learner \cite{van_der_laan_super_2007} in the case of K-fold cross validation. Its oracle property in model selection has been established both  asymptotically and in finite-sample \cite{van_der_laan_unified_2003}. That is,  on average, given the same set of estimated kernel predictors $\{\hat{h}_d\}_{d=1}^D$, the behavior of the ensemble made by the cross-validated selector converges in $O(\frac{1}{n})$ rate to the “oracle ensemble” made by an oracle that has access to infinite amount of validation data. Consequently, under the null hypothesis, given a set of base kernels $\{\hat{h}_d\}_{d=1}^D$ with diverse mathematical properties, the oracle selection property of CVEK guarantees the selection of a model ensemble that best describes the data, thereby resulting in correct Type I error by mitigating model misspecification under the null. 

\subsubsection{Generalization}

In limited samples, the oracle property does not ensure a powerful test, and the estimator's \textit{generalization} property, i.e. how fast the estimator's behavior approaches its asymptotic counterpart, must also be taken into consideration to guarantee good power under the alternative. This is because even under a oracle selector, an ensemble of flexible estimators in finite samples will still overfit the interaction effect under the alternative, leading to a test with low power. To explore this issue, we notice that CVEK's ensemble form corresponds to that of the \textit{Ensemble of Kernel Predictors} (EKP) \cite{cortes_ensembles_2011}, for which, under arbitrary choice of base kernels, the generalization error of the ensemble estimator converges at a rate of at least $O(\frac{1}{\sqrt{n}})$. However, for most epidemiological studies with small to moderate samples, a nonparametric rate of $O(\frac{1}{\sqrt{n}})$ may be too slow to merit powerful inference. 
It is therefore of great practical interest to understand if the ensemble's generalization performance can be improved beyond the rate of $O(\frac{1}{\sqrt{n}})$ with a careful selection of base kernels.

To this end, we draw upon a classical result from the RKHS literature that the generalization error of a kernel-based estimator is bounded above by its RKHS’s local Rademacher complexity \cite{bartlett_local_2005}, a measure of the richness of the class of candidate functions that is  characterized by the rate of eigenvalue decay of its  kernel function \cite{mendelson_performance_2003}. Consequently, since the learned ensemble estimator $\widehat{\bh} = \sum_{d=1}^D \widehat{u}_d \bh_d$ lies in convex combination of the RKHSs generated by the base kernels, the generalization property of the CVEK estimator is explicitly  characterized by the generalization property of the "selected" base estimators (i.e. the $\hat{\bh}_d$ assigned non-zero weight by the ensemble) in terms of their respective rates of eigenvalue decay. Moreover, it can be shown that for special classes of  kernels, the generalization error rate can indeed be improved upon: finite-rank kernels (e.g. linear and polynomial kernels) are able to achieve $O(\frac{1}{n})$, while kernel families with exponential rate eigenvalue decay (e.g. Gaussian RBF kernel) can achieve a rate of $O(\frac{log(n)}{n})$ \cite{cortes_learning_2013}.

\subsubsection{Choice for Base Kernels}
\label{sec:cvek_kernel}

In combination of the CVEK's oracle selection property, the above result suggests that, if there exists sets of parametric kernels in the library, CVEK behaves as a parametric model by achieving a rate  of $O(\frac{1}{n})$ if the data-generation function is indeed parametric. For more complex data-generation mechanism, CVEK is able to achieve a rate of $O(\frac{log(n)}{n})$ by including sets of Gaussian RBF kernels in the library. Consequently, we recommend practitioners to construct the kernel library with a mix of parametric kernels (linear, polynomial) and smooth kernels with exponential rate in eigendecay  (e.g. a collection of Gaussian RBF kernel with different fixed spatial smoothness parameters). We notice that despite the extreme smoothness of these base kernels, the resulting ensemble is in fact very flexible and hence does not risk underfitting the data generation function. This is because by Bochner's theorem, fitting the model with an ensemble of Gaussian RBF kernels is equivalent to approximating the spectral density of the function with convex mixtures of Gaussian densities, which is shown to be capable of approximating arbitrary continuous density with compact support in $\real^d$ \cite{bacharoglou_approximation_2010}. 
More generally, practitioners should be careful about using an ensemble containing only flexible kernels, since in order to represent a large function space, these kernel functions usually have heavy-tailed spectral densities and therefore slow eigenvalue decay \cite{mendelson_performance_2003}. Finally, we notice that there also exists interesting kernel families that lie outside the usual scope of theoretical analysis, but tend to do well in practice. One such example is the Neural Network kernel \cite{rasmussen_gaussian_2006}. We investigate the performance of these kernels and compare to the performance of the Gaussian RBF ensemble in Section \ref{sec:simu}.

\subsection{Detailed Simulation Results}
\label{sec:simu_detail}

In this section we document the value of estimated $\widehat{P}(p < 0.05)$ from the simulation presented in Section 6 (Simulation Experiment) of the paper. Recall that the simulation data is generated from below mechanism:
\begin{align*}
y_i &=  h_1(\bx_{i,1}) + h_2(\bx_{i,2}) + \delta * h_{12}(\bx_{i,1}, \bx_{i,2}) + \epsilon_i
\end{align*}
where $h_i$'s are functions with unit norm sampled from the reproducing kernel Hilbert spaces (RKHSs) generated by $k_{\texttt{true}}$, and the data is then fitted using Gaussian process with $k_{\texttt{model}}$. 

Each table documents the $\widehat{P}(p < 0.05)$ resulted from fixing $k_{\texttt{true}}$ to a Mat\'{e}rn kernel with specific value of smoothness parameter $\nu$ and complexity parameter $\sigma$, and then varying the strength of the interaction $\delta \in [0, 1]$ and the model kernel $k_{\texttt{model}}$. 

Our general observations are: 
\begin{enumerate}
\item The value of test power increases as the value $k_{\texttt{true}}$'s complexity parameter $\sigma$ becomes larger. This is possibly caused by the fact that the interaction becomes easier to detect as the pure interaction function $h_{12} \in \Hsc_{12}$ becomes more complex as in it varies more quickly.
\item Given the data-generation mechanism:
\begin{enumerate}
\item Polynomial kernels (Linear and Quadratic kernels) exhibits underfit, and result in inflated Type I error but also low power.
\item Lower-order Mat\'{e}rn kernels (Matern 1/2 and 3/2) tend to exhibits overfit for smoother $k_{\texttt{true}}$'s, and result in deflated Type I error and diminished low power. This conclusion cautions us against the approach of extending model complexity by naively relaxing model's smoothness (i.e. differentiability) constraint.
\item Gaussian RBF Kernels in general can perform well, but only if the hyperparameter is chosen carefully. Specifically, selecting the hyperparameter $\sigma$ by maximizing model likelihood does not perform well in small sample. On the other hand, the naive approach of selecting $\sigma$ by setting $\sigma$ to population median performs surprisingly well.
\item Neural Network kernels also work well in general. Their performance is also impacted by the hyperparameters. However not as sensitive as Gaussian RBF.
\end{enumerate}
\end{enumerate}

\begin{table}[!htbp] \centering 
  \caption{$k_{\texttt{true}}=$ Mat\'{e}rn 3/2, $\sigma = 0.5$} 
  \label{} 
\begin{tabular}{@{\extracolsep{5pt}} cccccccccc} 
\\[-1.8ex]\hline 
\hline \\[-1.8ex] 
$k_{\texttt{model}}  / \delta$ & 0 & 0.1 & 0.2 & 0.3 & 0.4 & 0.5 & 0.6 & 0.8 & 1 \\ 
\hline \\[-1.8ex] 
Linear & $0.194$ & $0.352$ & $0.444$ & $0.449$ & $0.521$ & $0$ & $0.485$ & $0.514$ & $0.473$ \\ 
Quadratic & $0.078$ & $0.326$ & $0.481$ & $0.588$ & $0.619$ & $0.653$ & $0.641$ & $0.653$ & $0.657$ \\ 
RBF\_MLE & $0.081$ & $0.400$ & $0.549$ & $0.529$ & $0.490$ & $0.473$ & $0.480$ & $0.484$ & $0.478$ \\ 
RBF\_Median & $0.038$ & $0.199$ & $0.453$ & $0.645$ & $0.695$ & $0.790$ & $0.782$ & $0.877$ & $0.889$ \\ 
Matern 1/2 & $0.020$ & $0.137$ & $0.296$ & $0.469$ & $0.539$ & $0.545$ & $0.555$ & $0.573$ & $0.604$ \\ 
Matern 3/2 & $0.047$ & $0.251$ & $0.471$ & $0.596$ & $0.674$ & $0.764$ & $0.783$ & $0.844$ & $0.872$ \\ 
Matern 5/2 & $0.035$ & $0.243$ & $0.458$ & $0.612$ & $0.688$ & $0.775$ & $0.833$ & $0.863$ & $0.896$ \\ 
NN 0.1 & $0.059$ & $0.299$ & $0.505$ & $0.563$ & $0.618$ & $0.640$ & $0.654$ & $0.666$ & $0.671$ \\ 
NN 1 & $0.047$ & $0.266$ & $0.504$ & $0.565$ & $0.655$ & $0.685$ & $0.732$ & $0.772$ & $0.805$ \\ 
NN 10 & $0.050$ & $0.234$ & $0.477$ & $0.631$ & $0.687$ & $0.769$ & $0.786$ & $0.874$ & $0.898$ \\ 
CVKE\_RBF & $0.044$ & $0.222$ & $0.441$ & $0.607$ & $0.682$ & $0.740$ & $0.792$ & $0.860$ & $0.893$ \\ 
CVKE\_NN & $0.041$ & $0.190$ & $0.405$ & $0.524$ & $0.622$ & $0.711$ & $0.758$ & $0.826$ & $0.844$ \\ 
\hline \\[-1.8ex] 
\end{tabular} 
\end{table}

\begin{table}[!htbp] \centering 
  \caption{$k_{\texttt{true}}=$ Mat\'{e}rn 3/2, $\sigma = 1$} 
  \label{} 
\begin{tabular}{@{\extracolsep{5pt}} cccccccccc} 
\\[-1.8ex]\hline 
\hline \\[-1.8ex] 
$k_{\texttt{model}}  / \delta$ & 0 & 0.1 & 0.2 & 0.3 & 0.4 & 0.5 & 0.6 & 0.8 & 1 \\ 
\hline \\[-1.8ex] 
Linear & $0.299$ & $0.481$ & $0.634$ & $0.696$ & $0.755$ & $0.716$ & $0.719$ & $0.739$ & $0.743$ \\ 
Quadratic & $0.113$ & $0.603$ & $0.726$ & $0.731$ & $0.749$ & $0.732$ & $0.774$ & $0.762$ & $0.744$ \\ 
RBF\_MLE & $0.174$ & $0.761$ & $0.876$ & $0.892$ & $0.874$ & $0.841$ & $0.825$ & $0.797$ & $0.804$ \\ 
RBF\_Median & $0.045$ & $0.556$ & $0.825$ & $0.893$ & $0.919$ & $0.948$ & $0.950$ & $0.961$ & $0.961$ \\ 
Matern 1/2 & $0.015$ & $0.272$ & $0.609$ & $0.748$ & $0.794$ & $0.818$ & $0.854$ & $0.873$ & $0.877$ \\ 
Matern 3/2 & $0.044$ & $0.574$ & $0.808$ & $0.896$ & $0.914$ & $0.935$ & $0.936$ & $0.949$ & $0.933$ \\ 
Matern 5/2 & $0.040$ & $0.606$ & $0.807$ & $0.873$ & $0.854$ & $0.874$ & $0.904$ & $0.908$ & $0.886$ \\ 
NN 0.1 & $0.081$ & $0.593$ & $0.718$ & $0.718$ & $0.721$ & $0.752$ & $0.733$ & $0.750$ & $0.758$ \\ 
NN 1 & $0.058$ & $0.608$ & $0.752$ & $0.761$ & $0.775$ & $0.755$ & $0.771$ & $0.759$ & $0.787$ \\ 
NN 10 & $0.046$ & $0.578$ & $0.848$ & $0.880$ & $0.913$ & $0.919$ & $0.913$ & $0.929$ & $0.912$ \\ 
CVKE\_RBF & $0.041$ & $0.578$ & $0.811$ & $0.881$ & $0.912$ & $0.938$ & $0.951$ & $0.965$ & $0.949$ \\ 
CVKE\_NN & $0.032$ & $0.541$ & $0.738$ & $0.854$ & $0.911$ & $0.927$ & $0.939$ & $0.948$ & $0.947$ \\ 
\hline \\[-1.8ex] 
\end{tabular} 
\end{table} 

\begin{table}[!htbp] \centering 
  \caption{$k_{\texttt{true}}=$ Mat\'{e}rn 3/2, $\sigma = 1.5$} 
  \label{} 
\begin{tabular}{@{\extracolsep{5pt}} cccccccccc} 
\\[-1.8ex]\hline 
\hline \\[-1.8ex] 
$k_{\texttt{model}}  / \delta$ & 0 & 0.1 & 0.2 & 0.3 & 0.4 & 0.5 & 0.6 & 0.8 & 1 \\ 
\hline \\[-1.8ex] 
Linear & $0.299$ & $0.457$ & $0.655$ & $0.701$ & $0.755$ & $0.772$ & $0.792$ & $0.785$ & $0.810$ \\ 
Quadratic & $0.123$ & $0.619$ & $0.756$ & $0.829$ & $0.806$ & $0.788$ & $0.815$ & $0.812$ & $0.830$ \\ 
RBF\_MLE & $0.239$ & $0.822$ & $0.916$ & $0.932$ & $0.928$ & $0.912$ & $0.895$ & $0.882$ & $0.861$ \\ 
RBF\_Median & $0.040$ & $0.676$ & $0.881$ & $0.913$ & $0.947$ & $0.955$ & $0.969$ & $0.957$ & $0.936$ \\ 
Matern 1/2 & $0.010$ & $0.282$ & $0.667$ & $0.802$ & $0.858$ & $0.872$ & $0.880$ & $0.894$ & $0.916$ \\ 
Matern 3/2 & $0.043$ & $0.675$ & $0.880$ & $0.941$ & $0.950$ & $0.942$ & $0.941$ & $0.952$ & $0.943$ \\ 
Matern 5/2 & $0.041$ & $0.678$ & $0.883$ & $0.923$ & $0.908$ & $0.903$ & $0.909$ & $0.896$ & $0.890$ \\ 
NN 0.1 & $0.073$ & $0.671$ & $0.785$ & $0.801$ & $0.788$ & $0.817$ & $0.797$ & $0.822$ & $0.806$ \\ 
NN 1 & $0.046$ & $0.703$ & $0.806$ & $0.817$ & $0.811$ & $0.817$ & $0.815$ & $0.790$ & $0.828$ \\ 
NN 10 & $0.031$ & $0.702$ & $0.881$ & $0.933$ & $0.919$ & $0.922$ & $0.910$ & $0.911$ & $0.915$ \\ 
CVKE\_RBF & $0.042$ & $0.681$ & $0.860$ & $0.930$ & $0.945$ & $0.947$ & $0.946$ & $0.960$ & $0.947$ \\ 
CVKE\_NN & $0.034$ & $0.650$ & $0.863$ & $0.895$ & $0.942$ & $0.941$ & $0.944$ & $0.944$ & $0.946$ \\ 
\hline \\[-1.8ex] 
\end{tabular} 
\end{table} 

\begin{table}[!htbp] \centering 
  \caption{$k_{\texttt{true}}=$ Mat\'{e}rn 5/2, $\sigma = 0.5$} 
  \label{} 
\begin{tabular}{@{\extracolsep{5pt}} cccccccccc} 
\\[-1.8ex]\hline 
\hline \\[-1.8ex] 
$k_{\texttt{model}}  / \delta$ & 0 & 0.1 & 0.2 & 0.3 & 0.4 & 0.5 & 0.6 & 0.8 & 1 \\ 
\hline \\[-1.8ex] 
Linear & $0.174$ & $0.351$ & $0.507$ & $0.539$ & $0.533$ & $0.559$ & $0.552$ & $0.579$ & $0.553$ \\ 
Quadratic & $0.055$ & $0.107$ & $0.186$ & $0.253$ & $0.284$ & $0.359$ & $0.394$ & $0.508$ & $0.563$ \\ 
RBF\_MLE & $0.061$ & $0.137$ & $0.174$ & $0.209$ & $0.270$ & $0.361$ & $0.357$ & $0.411$ & $0.468$ \\ 
RBF\_Median & $0.052$ & $0.091$ & $0.162$ & $0.214$ & $0.250$ & $0.306$ & $0.323$ & $0.392$ & $0.445$ \\ 
Matern 1/2 & $0.015$ & $0.058$ & $0.092$ & $0.140$ & $0.175$ & $0.176$ & $0.190$ & $0.201$ & $0.218$ \\ 
Matern 3/2 & $0.041$ & $0.089$ & $0.148$ & $0.203$ & $0.242$ & $0.283$ & $0.300$ & $0.348$ & $0.421$ \\ 
Matern 5/2 & $0.056$ & $0.099$ & $0.154$ & $0.222$ & $0.275$ & $0.323$ & $0.345$ & $0.433$ & $0.519$ \\ 
NN 0.1 & $0.059$ & $0.111$ & $0.178$ & $0.235$ & $0.277$ & $0.332$ & $0.365$ & $0.480$ & $0.498$ \\ 
NN 1 & $0.038$ & $0.091$ & $0.161$ & $0.224$ & $0.281$ & $0.332$ & $0.380$ & $0.455$ & $0.522$ \\ 
NN 10 & $0.039$ & $0.113$ & $0.165$ & $0.213$ & $0.271$ & $0.304$ & $0.339$ & $0.418$ & $0.476$ \\ 
CVKE\_RBF & $0.049$ & $0.083$ & $0.155$ & $0.221$ & $0.279$ & $0.339$ & $0.435$ & $0.509$ & $0.586$ \\ 
CVKE\_NN & $0.039$ & $0.085$ & $0.186$ & $0.245$ & $0.295$ & $0.306$ & $0.377$ & $0.436$ & $0.549$ \\ 
\hline \\[-1.8ex] 
\end{tabular} 
\end{table} 

\begin{table}[!htbp] \centering 
  \caption{$k_{\texttt{true}}=$ Mat\'{e}rn 5/2, $\sigma = 1$} 
  \label{} 
\begin{tabular}{@{\extracolsep{5pt}} cccccccccc} 
\\[-1.8ex]\hline 
\hline \\[-1.8ex] 
$k_{\texttt{model}}  / \delta$ & 0 & 0.1 & 0.2 & 0.3 & 0.4 & 0.5 & 0.6 & 0.8 & 1 \\ 
\hline \\[-1.8ex] 
Linear & $0.229$ & $0.396$ & $0.471$ & $0.517$ & $0.515$ & $0.523$ & $0.528$ & $0.531$ & $0.518$ \\ 
Quadratic & $0.071$ & $0.333$ & $0.517$ & $0.654$ & $0.703$ & $0.801$ & $0.793$ & $0.825$ & $0.869$ \\ 
RBF\_MLE & $0.077$ & $0.313$ & $0.489$ & $0.558$ & $0.574$ & $0.621$ & $0.616$ & $0.619$ & $0.544$ \\ 
RBF\_Median & $0.050$ & $0.251$ & $0.455$ & $0.576$ & $0.648$ & $0.729$ & $0.767$ & $0.840$ & $0.883$ \\ 
Matern 1/2 & $0.012$ & $0.089$ & $0.292$ & $0.430$ & $0.457$ & $0.477$ & $0.543$ & $0.568$ & $0.565$ \\ 
Matern 3/2 & $0.039$ & $0.230$ & $0.444$ & $0.584$ & $0.657$ & $0.748$ & $0.761$ & $0.822$ & $0.863$ \\ 
Matern 5/2 & $0.052$ & $0.287$ & $0.475$ & $0.636$ & $0.692$ & $0.770$ & $0.823$ & $0.842$ & $0.896$ \\ 
NN 0.1 & $0.059$ & $0.303$ & $0.531$ & $0.626$ & $0.691$ & $0.780$ & $0.799$ & $0.847$ & $0.867$ \\ 
NN 1 & $0.052$ & $0.292$ & $0.508$ & $0.645$ & $0.708$ & $0.763$ & $0.785$ & $0.874$ & $0.865$ \\ 
NN 10 & $0.043$ & $0.299$ & $0.493$ & $0.624$ & $0.693$ & $0.785$ & $0.787$ & $0.860$ & $0.869$ \\ 
CVKE\_RBF & $0.037$ & $0.263$ & $0.470$ & $0.623$ & $0.710$ & $0.786$ & $0.795$ & $0.869$ & $0.895$ \\ 
CVKE\_NN & $0.049$ & $0.237$ & $0.449$ & $0.563$ & $0.660$ & $0.694$ & $0.771$ & $0.835$ & $0.871$ \\ 
\hline \\[-1.8ex] 
\end{tabular} 
\end{table} 

\begin{table}[!htbp] \centering 
  \caption{$k_{\texttt{true}}=$ Mat\'{e}rn 5/2, $\sigma = 1.5$} 
  \label{} 
\begin{tabular}{@{\extracolsep{5pt}} cccccccccc} 
\\[-1.8ex]\hline 
\hline \\[-1.8ex] 
$k_{\texttt{model}}  / \delta$ & 0 & 0.1 & 0.2 & 0.3 & 0.4 & 0.5 & 0.6 & 0.8 & 1 \\ 
\hline \\[-1.8ex] 
Linear & $0.343$ & $0.561$ & $0.724$ & $0.782$ & $0.830$ & $0.800$ & $0.821$ & $0.813$ & $0.821$ \\ 
Quadratic & $0.096$ & $0.723$ & $0.840$ & $0.875$ & $0.881$ & $0.881$ & $0.908$ & $0.911$ & $0.885$ \\ 
RBF\_MLE & $0.082$ & $0.743$ & $0.899$ & $0.911$ & $0.905$ & $0.876$ & $0.886$ & $0.854$ & $0.836$ \\ 
RBF\_Median & $0.038$ & $0.684$ & $0.858$ & $0.935$ & $0.952$ & $0.954$ & $0.956$ & $0.961$ & $0.962$ \\ 
Matern 1/2 & $0.016$ & $0.360$ & $0.663$ & $0.802$ & $0.846$ & $0.883$ & $0.879$ & $0.896$ & $0.896$ \\ 
Matern 3/2 & $0.034$ & $0.698$ & $0.853$ & $0.925$ & $0.944$ & $0.952$ & $0.941$ & $0.968$ & $0.971$ \\ 
Matern 5/2 & $0.046$ & $0.733$ & $0.877$ & $0.921$ & $0.930$ & $0.939$ & $0.954$ & $0.942$ & $0.949$ \\ 
NN 0.1 & $0.059$ & $0.721$ & $0.837$ & $0.856$ & $0.875$ & $0.890$ & $0.903$ & $0.900$ & $0.903$ \\ 
NN 1 & $0.039$ & $0.700$ & $0.870$ & $0.897$ & $0.865$ & $0.899$ & $0.904$ & $0.917$ & $0.916$ \\ 
NN 10 & $0.044$ & $0.729$ & $0.888$ & $0.928$ & $0.920$ & $0.953$ & $0.948$ & $0.960$ & $0.946$ \\ 
CVKE\_RBF & $0.031$ & $0.708$ & $0.887$ & $0.928$ & $0.940$ & $0.947$ & $0.948$ & $0.960$ & $0.957$ \\ 
CVKE\_NN & $0.032$ & $0.671$ & $0.859$ & $0.925$ & $0.935$ & $0.949$ & $0.946$ & $0.954$ & $0.966$ \\ 
\hline \\[-1.8ex] 
\end{tabular} 
\end{table}


\begin{table}[!htbp] \centering 
  \caption{$k_{\texttt{true}}=$ Gaussian RBF, $\sigma = 0.5$} 
  \label{} 
\begin{tabular}{@{\extracolsep{5pt}} cccccccccc} 
\\[-1.8ex]\hline 
\hline \\[-1.8ex] 
$k_{\texttt{model}}  / \delta$ & 0 & 0.1 & 0.2 & 0.3 & 0.4 & 0.5 & 0.6 & 0.8 & 1 \\ 
\hline \\[-1.8ex] 
Linear & $0.143$ & $0.341$ & $0.484$ & $0.537$ & $0.585$ & $0.559$ & $0.574$ & $0.576$ & $0.560$ \\ 
Quadratic & $0.069$ & $0.128$ & $0.206$ & $0.273$ & $0.355$ & $0.412$ & $0.443$ & $0.545$ & $0.620$ \\ 
RBF\_MLE & $0.068$ & $0.148$ & $0.226$ & $0.263$ & $0.305$ & $0.377$ & $0.429$ & $0.459$ & $0.479$ \\ 
RBF\_Median & $0.045$ & $0.100$ & $0.181$ & $0.245$ & $0.318$ & $0.343$ & $0.354$ & $0.473$ & $0.533$ \\ 
Matern 1/2 & $0.029$ & $0.059$ & $0.113$ & $0.167$ & $0.198$ & $0.199$ & $0.245$ & $0.251$ & $0.254$ \\ 
Matern 3/2 & $0.045$ & $0.092$ & $0.171$ & $0.247$ & $0.286$ & $0.320$ & $0.361$ & $0.458$ & $0.472$ \\ 
Matern 5/2 & $0.046$ & $0.124$ & $0.181$ & $0.271$ & $0.319$ & $0.403$ & $0.427$ & $0.495$ & $0.561$ \\ 
NN 0.1 & $0.054$ & $0.118$ & $0.194$ & $0.268$ & $0.345$ & $0.375$ & $0.451$ & $0.515$ & $0.593$ \\ 
NN 1 & $0.055$ & $0.118$ & $0.194$ & $0.287$ & $0.322$ & $0.379$ & $0.402$ & $0.513$ & $0.574$ \\ 
NN 10 & $0.042$ & $0.103$ & $0.184$ & $0.239$ & $0.335$ & $0.348$ & $0.407$ & $0.482$ & $0.517$ \\ 
CVKE\_RBF & $0.041$ & $0.103$ & $0.215$ & $0.323$ & $0.315$ & $0.414$ & $0.486$ & $0.601$ & $0.679$ \\ 
CVKE\_NN & $0.044$ & $0.117$ & $0.157$ & $0.301$ & $0.330$ & $0.411$ & $0.477$ & $0.538$ & $0.616$ \\ 
\hline \\[-1.8ex] 
\end{tabular} 
\end{table} 

\begin{table}[!htbp] \centering 
  \caption{$k_{\texttt{true}}=$ Gaussian RBF, $\sigma = 1$} 
  \label{} 
\begin{tabular}{@{\extracolsep{5pt}} cccccccccc} 
\\[-1.8ex]\hline 
\hline \\[-1.8ex] 
$k_{\texttt{model}}  / \delta$ & 0 & 0.1 & 0.2 & 0.3 & 0.4 & 0.5 & 0.6 & 0.8 & 1 \\ 
\hline \\[-1.8ex] 
Linear & $0.286$ & $0.396$ & $0.457$ & $0.525$ & $0.520$ & $0.536$ & $0.527$ & $0.521$ & $0.555$ \\ 
Quadratic & $0.056$ & $0.203$ & $0.369$ & $0.490$ & $0.546$ & $0.658$ & $0.702$ & $0.783$ & $0.844$ \\ 
RBF\_MLE & $0.065$ & $0.234$ & $0.330$ & $0.430$ & $0.507$ & $0.554$ & $0.577$ & $0.608$ & $0.601$ \\ 
RBF\_Median & $0.046$ & $0.161$ & $0.297$ & $0.421$ & $0.502$ & $0.570$ & $0.587$ & $0.693$ & $0.772$ \\ 
Matern 1/2 & $0.016$ & $0.068$ & $0.183$ & $0.247$ & $0.273$ & $0.320$ & $0.361$ & $0.394$ & $0.424$ \\ 
Matern 3/2 & $0.042$ & $0.198$ & $0.307$ & $0.433$ & $0.504$ & $0.558$ & $0.588$ & $0.670$ & $0.764$ \\ 
Matern 5/2 & $0.043$ & $0.184$ & $0.340$ & $0.458$ & $0.510$ & $0.607$ & $0.655$ & $0.720$ & $0.789$ \\ 
NN 0.1 & $0.053$ & $0.216$ & $0.373$ & $0.456$ & $0.552$ & $0.639$ & $0.670$ & $0.770$ & $0.836$ \\ 
NN 1 & $0.045$ & $0.185$ & $0.354$ & $0.481$ & $0.545$ & $0.647$ & $0.678$ & $0.788$ & $0.808$ \\ 
NN 10 & $0.044$ & $0.175$ & $0.347$ & $0.465$ & $0.510$ & $0.579$ & $0.664$ & $0.730$ & $0.763$ \\ 
CVKE\_RBF & $0.043$ & $0.193$ & $0.346$ & $0.452$ & $0.533$ & $0.633$ & $0.690$ & $0.801$ & $0.870$ \\ 
CVKE\_NN & $0.043$ & $0.162$ & $0.318$ & $0.467$ & $0.552$ & $0.671$ & $0.696$ & $0.772$ & $0.834$ \\ 
\hline \\[-1.8ex] 
\end{tabular} 
\end{table} 

\begin{table}[!htbp] \centering 
  \caption{$k_{\texttt{true}}=$ Gaussian RBF, $\sigma = 1.5$} 
  \label{} 
\begin{tabular}{@{\extracolsep{5pt}} cccccccccc} 
\\[-1.8ex]\hline 
\hline \\[-1.8ex] 
$k_{\texttt{model}} / \delta$ & 0 & 0.1 & 0.2 & 0.3 & 0.4 & 0.5 & 0.6 & 0.8 & 1 \\ 
\hline \\[-1.8ex] 
Linear & $0.347$ & $0.471$ & $0.569$ & $0.625$ & $0.660$ & $0.646$ & $0.640$ & $0.608$ & $0.662$ \\ 
Quadratic & $0.080$ & $0.554$ & $0.767$ & $0.854$ & $0.883$ & $0.913$ & $0.922$ & $0.941$ & $0.956$ \\ 
RBF\_MLE & $0.052$ & $0.555$ & $0.755$ & $0.804$ & $0.840$ & $0.819$ & $0.792$ & $0.766$ & $0.712$ \\ 
RBF\_Median & $0.046$ & $0.481$ & $0.719$ & $0.795$ & $0.882$ & $0.902$ & $0.914$ & $0.950$ & $0.946$ \\ 
Matern 1/2 & $0.014$ & $0.218$ & $0.482$ & $0.591$ & $0.673$ & $0.704$ & $0.765$ & $0.756$ & $0.782$ \\ 
Matern 3/2 & $0.036$ & $0.494$ & $0.686$ & $0.825$ & $0.862$ & $0.903$ & $0.920$ & $0.943$ & $0.939$ \\ 
Matern 5/2 & $0.047$ & $0.543$ & $0.755$ & $0.829$ & $0.869$ & $0.903$ & $0.925$ & $0.946$ & $0.955$ \\ 
NN 0.1 & $0.054$ & $0.581$ & $0.774$ & $0.866$ & $0.884$ & $0.919$ & $0.929$ & $0.946$ & $0.968$ \\ 
NN 1 & $0.037$ & $0.553$ & $0.750$ & $0.834$ & $0.891$ & $0.923$ & $0.926$ & $0.947$ & $0.969$ \\ 
NN 10 & $0.044$ & $0.523$ & $0.756$ & $0.827$ & $0.877$ & $0.901$ & $0.922$ & $0.956$ & $0.950$ \\ 
CVKE\_RBF & $0.034$ & $0.554$ & $0.741$ & $0.855$ & $0.877$ & $0.921$ & $0.946$ & $0.961$ & $0.979$ \\ 
CVKE\_NN & $0.043$ & $0.534$ & $0.749$ & $0.855$ & $0.877$ & $0.936$ & $0.939$ & $0.950$ & $0.954$ \\ 
\hline \\[-1.8ex] 
\end{tabular} 
\end{table} 

\end{document}